\documentclass[11pt,a4paper,cits]{article}
\usepackage{jheppub}
\def\unit{\relax{\rm 1\kern-.26em I}}

\def\bea{\begin{eqnarray}}
\def\eea{\end{eqnarray}}
\def\be{\begin{equation}}
\def\ee{\end{equation}}
\def\nn{\nonumber}

\newcommand{\gsim}{\lower.7ex\hbox{$\;\stackrel{\textstyle>}{\sim}\;$}}
\newcommand{\lsim}{\lower.7ex\hbox{$\;\stackrel{\textstyle<}{\sim}\;$}}

\title{On the Effective Description of Large Volume Compactifications}
\author{Diego Gallego}
\affiliation{Escuela de F\'isica, Universidad Pedag\'ogica y
Tecnol\'ogica de Colombia (UPTC),\\ Tunja, Colombia}
\emailAdd{diego.gallego@uptc.edu.co}

\abstract{
We study the reliability of the Two-Step moduli stabilization in
the type-IIB Large Volume Scenarios with matter and gauge
interactions. The general analysis is based on a family of ${\cal
N}=1$ Supergravity models with a factorizable K\"ahler invariant
function, where the decoupling between two sets of fields without
a mass hierarchy is easily understood. For the Large Volume
Scenario particular analyses are performed for explicit models,
one of such developed for the first time here, finding that the
simplified version, where the Dilaton and Complex structure moduli
are regarded as frozen by a previous stabilization, is a reliable
supersymmetric description whenever the neglected fields stand at
their leading $F$-flatness conditions and be neutral. The terms
missed by the simplified approach are either suppressed by powers
of the Calabi-Yau volume, or are higher order operators in the
matter fields, and then irrelevant for the moduli stabilization
procedure. Although the power of the volume suppressing such
corrections depends on the particular model, up to the mass level
it is independent of the modular weight for the matter fields.
This at least for the models studied here but we give arguments to
expect the same in general. These claims are checked through
numerical examples.
\\
We discuss how the factorizable models present a context where
despite the lack of a hierarchy with the supersymmetry breaking
scale, the effective theory still has a supersymmetric
description. This can be understood from the fact that it is
possible to find vanishing solution for the auxiliary components
of the fields being integrated out, independently of the remaining
dynamics.
\\
Our results settle down the question on the reliability of the way
the Dilaton and Complex structure are treated in type-IIB
compactifications with large compact manifold volumina.}


\begin{document}
\maketitle

\section{Introduction}
The great progress that string theory phenomenology has achieved
in recent years is in contrast with the several approximations
these studies have always implicit, being one of the most drastic
to neglect a large subset of fields. The first candidates to be
neglected are the string excitations and Kaluza-Klein (KK) modes
which are always very heavy fields, with masses dictated by the
string and compactification scales. Then, decoupling
\cite{Appelquist:1974tg} and symmetry arguments can be used to
argue for an ``effective'' theory, although these have not been
properly integrated out \cite{Witten:1985xb}. At this level the
truncation is fine since these fields and their couplings are easy
to be identified. Nevertheless, the Supergravity (SUGRA) theory
describing the remaining light modes is still very complicated
such that an explicit and precise study continues to be extremely
difficult. A further truncation is then compulsory but now over
fields whose masses and couplings are highly sensitive to the
point we stand in the field space, so that this procedure starts
to be less clear.
\\
This is precisely the philosophy adopted in the stabilization of
moduli fields in string compactifications. In this context the
moduli to be fixed by tree level effects in the superpotential,
flux induced terms \cite{Giddings:2001yu} which are quantized and
therefore naturally of order one in Planck units, are regarded as
completely frozen. Then the rest of the fields are stabilized
using non-perturbative effects, that are naturally suppressed. As
presented this Two-Step procedure seems alright due to the
hierarchy between the sources of the dynamics stabilizing each
sector; however, the moduli couplings and even their masses are
not as easy to spot as in the case of the string and KK modes,
then a more careful study is required. In recent years several
studies has tackled this issue from different perspectives (see
for example
\cite{deAlwis:2005tf,deAlwis:2005tg,Achucarro:2007qa,Choi:2008hn,Achucarro:2008sy,Achucarro:2008fk,Gallego:2008qi,Choi:2009jn,Gallego:2009px,Brizi:2009nn,Achucarro:2010da,Brizi:2010ab,Achucarro:2010jv}).
Particularly in \cite{Gallego:2008qi,Gallego:2009px} things were
settled down for models where the perturbative part of the
superpotential takes a tiny value at the vacuum. There it was
pointed out that this tuning on the superpotential, firstly
motivated in the seminal work of Kachru-Kallosh-Linde-Trivedi
(KKLT) \cite{Kachru:2003aw} as such to obtain good
phenomenological Vacuum Expectation Values (VEV) for the moduli,
is also a necessary requirement for the reliability of the
Two-Step procedure, as it ensures a mass hierarchy between the two
sets of fields. These results apply to the several sequels of KKLT
where the VEV of the superpotential is quite small, and the fields
to be fixed by the fluxes are completely disregarded focusing only
on the second step where the remaining fields are stabilized.
\\
An important point here is the fact that the theory used in the
second step is still a ${\cal N}=1$ SUGRA theory, so the first
step is supposed to proceed in a supersymmetric (SUSY) way. From
this point of view the tuning on the superpotential can be
understood using the fact that the SUSY breaking scale in these
KKLT scenarios is dictated mainly by this VEV, so that being small
the heavy fields decouple from SUSY breaking effects and the
effective theory is approximately SUSY \cite{Brizi:2009nn}.
\\
Following this logic, things seems more subtle for models where
the neglected fields acquire masses that are of the same order of
the masses of the fields to be stabilized and the SUSY breaking
scale. In this class of models fall the Large Volume Scenarios
(LVS) \cite{Balasubramanian:2005zx} where the masses and scales
are mainly ruled by the size of the compact manifold developing
exponential sized volume.
\\
The LVS has become one of the paradigms of string moduli
stabilization being widely studied (see for instance
\cite{Conlon:2005ki,Conlon:2006wz,Cremades:2007ig,Conlon:2008cj,Cicoli:2008va}
and more recent progress in
\cite{Conlon:2008wa,Anguelova:2009ht,Conlon:2010ji,Conlon:2010jq,Cicoli:2010yj,Krippendorf:2009zza}).
The main appealing feature of this scenario is the fact that it is
not necessary to claim for a tuned VEV for the superpotential in
order to obtain a low SUSY breaking scale. However, as usual in
any moduli stabilization scenario in type-IIB orientifold
compactification, the Dilaton and Complex structure moduli are
regarded as frozen although their masses lie on the same scale off
the remaining field masses, as can be understood from the SUGRA
contributions to the masses, coming like $m^2\sim \langle
e^K|W|^2\rangle$, with $K$ the K\"ahler potential and $W$ the
superpotential. Still, from the structure of the scalar potential,
in the large volume limit, it is possible to argue for a
decoupling \cite{Balasubramanian:2005zx} and systematic treatments
on this issue have been worked out
\cite{Achucarro:2008sy,Gallego:2008qi}. The idea in this cases is
the realization of a weak coupling between the two sectors so that
the stabilization of one side be almost insensitive to the other
one.
\\
In absence of gauge interactions such a decoupling is achieved if
the system with two sectors, identified by $H$, the ``heavy''
fields, and $L$, the remaining ones in the effective theory, is
described by a K\"ahler invariant function,
$G=K+M_P^2\log\left(|W|^2/M_P^6\right)$, $M_P$ the reduced Planck
mass, with a factorizable form, i.e.,
\begin{equation}
G(H,\bar H, L,\bar L)=G_H(H,\bar H)+G_L(L,\bar L)\,.
\end{equation}
This scenario was proposed by Binetruy et al.
\cite{Binetruy:2004hh} and extensively studied by Achucarro et al.
in several works
\cite{Achucarro:2007qa,Achucarro:2008sy,Achucarro:2008fk,Achucarro:2010da,Achucarro:2010jv}.
In \cite{Achucarro:2008sy,Gallego:2008qi} it was pointed out that
a Two-Step procedure in the LVS scenario, for the pure moduli
case, can be validated through the same generic arguments since
all mixing terms in $G$ between the K\"ahler sector and the
Dilaton and/or the Complex structure moduli are suppressed either
by powers of the volume or by non-perturbative effects. Despite
few comments in \cite{Achucarro:2008fk} these studies where
restricted to the case where no gauge interactions nor matter
fields were involved. Remains, then, and is the aim of this paper,
a systematic study for these more realistic scenarios.
\\
We will follow the same philosophy of
\cite{Gallego:2008qi,Gallego:2009px} by performing a proper
integration of the $H$ fields and compare the resulting theory
with the simple naive theory were the $H$ fields are simply
frozen. Recently Brizi et al. studied the way fields can be
integrated out in a SUSY fashion for SUGRA theories
\cite{Brizi:2009nn}. One of their results is that as far as there
is a hierarchy between the mass of the fields integrated out and
the SUSY breaking scale, one can neglect these breaking effects
and integrate the fields in a SUSY way, so that the effective
theory, at first approximation in a derivative expansion, is still
a SUGRA theory. They, moreover, found that the integration of
chiral multiplets proceeds, again at first order, through a chiral
equation of motion (e.o.m.) which coincides with the usual flat
expression, $\partial_H W=0$. In our situation this cannot be the
case as the neglected terms come like $ W$, which we suppose
naturally large, an therefore no longer negligible. However, if
the Two-Step procedure is approximately right, the effective
theory once the $H$ fields are integrated out should be also
approximately SUSY. Then, despite the lack of a chiral e.o.m. that
can be exploited in order to make the procedure fully SUSY
manifest, as was done in \cite{Gallego:2009px}, we follow a
manifestly SUSY procedure by keeping the auxiliary fields in the
Lagrangian and integrating out simultaneously all components of
the $H$ chiral multiplets. This not only leads to a
straightforward identification of the effective theory as an
approximate SUGRA theory but also turns out to simplify the
integration of the $H$ multiplets, at least in a schematic form,
in the case where gauge interactions are involved. The fact that
for the factorizable models the effective theory continues to be a
SUGRA one, at least approximately, then can be understood from the
fact that the weak coupling between the two sectors allows to have
approximately vanishing solutions for the $H$ auxiliary fields,
independently of the remaining dynamics. These models present,
therefore, an exception to the general case studied in
\cite{Brizi:2009nn} where is the hierarchy between the scales what
suppresses the SUSY breaking effects from the $H$ sector, but we
leave a fully superfield understanding of this situation for a
forthcoming paper \cite{Inprep}.
\\
The first thing to be noticed, even before turning on the gauge
interactions, is that the presence of matter fields clashes with
the factorizability of the system as their wave functions in
general depend on fields from both sectors and a large mixing can
be realized. However, if the wave function somehow turns out to be
suppressed, the mixing between the moduli is still safe, and
moreover the dynamics related to the $Q$ multiplets is suppressed
such that the independence of the $H$ solutions on the other
fields continues to hold. In the LVS this suppression is indeed
realized, with the wave functions coming like $1/{\cal V}^n$, with
the modular weight $n$ a positive number and ${\cal V}$ the volume
of the compact manifold \cite{Conlon:2006tj}. In general one can
expect, then, that the corrections to the simplified model be
modular weight dependent and the reliability of the Two-Step
procedure be constrained by this number. Nicely enough, due to a
necessary tuning on the numerical parameters, and the way the
matter fields stabilization proceeds, the corrections up to the
mass level are not only suppressed but also independent of $n$!.
Since for moduli stabilization issues this order in the
fluctuations is enough, this universality on the corrections is an
appealing feature of the LVS as a robust playground for moduli
stabilization models.
\\
Notice that we are being cavalier using the term effective theory:
in case there is a mass hierarchy we could properly speak about an
effective theory in the Wilson sense, but in our case things are
different and what we call effective theory is simply the one
obtained by solving the classical e.o.m. and then plugging back
the solution on the original Lagrangian, resulting with what we
will call effective Lagrangian. Nevertheless, the approximate
factorizability of the K\"ahler invariant function implies also
decoupling of the wave functions, indeed the scalar manifold is at
first order factorizable, so that a two derivative approximation
is still valid once one requires slow varying solutions in the $H$
sector. This is: the fluctuations in the $L$ sector do not excite,
at first order, fluctuations in the $H$ sector.
\\
The introduction of matter fields and gauge interactions in the
LVS is far from being just and academic exercise. Indeed, the
simplest pure moduli realization of LVS turns out to realize only
a deep AdS vacuum, therefore extra ingredients are needed to
uplift the vacuum being the more natural ones adding extra
fields.\footnote{Like in the KKLT scenario \cite{Kachru:2003aw} it
is possible to introduce $\overline{D3}$-branes at the tip of a
throat to do the job. This necessarily introduces explicit SUSY
breaking terms which one might want to avoid.} Although developing
models realizing exponentially sized compact manifold volumina is
not as prolific as the ones presenting moduli stabilization {\it
\`a la} KKLT, mainly due to the rather intricate structure leading
the stabilization which makes difficult its generalization, there
are a couple of proposals leading to Minkowski vacua, both of them
with matter and gauge interactions
\cite{Cremades:2007ig,Krippendorf:2009zza}.
\\
The structure of the paper is as follows: section
\ref{ONLYCHIRALSECT} is devoted to generalize the definition of
nearly factorizable models with the outlook of introducing matter
like fields. With this generalized setup, but without gauge
interactions, we show how the decoupling is realized and freezing
is a reliable procedure at leading order in a small parameter
characterizing the mixing in the K\"ahler invariant function. The
analysis is then particularized to the LVS where, however, there
are in principle several unrelated suppression factors for the
mixing, namely inverse powers of the volume in the K\"ahler
potential, and non-perturbative effects in the superpotential.
Besides of recasting and generalizing some of the results of
\cite{Achucarro:2008sy,Gallego:2008qi} the study is also performed
in a SUSY manifest way that helps to spot easily the SUSY nature
of the effective theory and moreover shows itself powerful when
dealing with gauge interactions; the main target of the paper lies
in section \ref{VectorAndChiralSect} where gauge dynamics are
introduced in the study, with the outcome of a constrain on the
gauge kinetic function for the general setup in order the Two-Step
procedure be reliable. The analysis is then performed for several
LVS models where it is shown that the restriction on the gauge
kinetic function is naturally avoided. The section closes with few
comments about higher order operators and its reliability in the
simplified model. Also few arguments are given in support of the
independence of the corrections on the modular weights; section
\ref{ConcluSect} is devoted for the conclusions and some
discussion; two appendices are left. One to review the main
aspects of the LVS models needed for the study, and to introduce a
novel model developing LVS. The second one gives numerical samples
showing explicitly the results obtained analytically in the main
text.

\section{Only chiral multiplets}\label{ONLYCHIRALSECT}


\subsection{Nearly factorizable models}

For our purposes it will be convenient to work in a K\"ahler gauge
where the factorizable nature of the ${\cal N}=1$ SUGRA theory be
manifested
\cite{Achucarro:2008sy,Achucarro:2008fk,Binetruy:2004hh}. This
leads us to work directly with the generalized K\"ahler invariant
function $G=K+M_P^2\log\left(\frac{|W|^2}{M_P^6}\right)$, where
$W$ is the superpotential, $K$ the K\"ahler potential and $M_P$
the reduced Planck mass that for simplicity we set to one in the
following. The approximate factorizable models are defined as such
that only suppressed terms mix two sectors identified by $H$ and
$L$, i.e.,
\begin{equation}\label{NEARFACTG}
G(H,\bar H,L,\bar L)=G_H(H,\bar H)+G_L(L,\bar L)+\epsilon
G_{mix}(H,\bar H, L,\bar  H)\,,
\end{equation}
with $\epsilon$ a small parameter. This form was proposed by
Binetruy et al. in \cite{Binetruy:2004hh}, to describe a SUSY
decoupling between the two sectors of fields; however, it lacks
from including matter like superfields. Indeed, although the form
for $G$ can be justified between moduli sectors, closed string
fields, it is not once matter, open string fields, enter in the
game since their wave functions depend on both moduli sectors, so
factorizability is lost once these fields acquire ${\cal O}(1)$
VEV. This situation can be easily taken into account generalizing
the factorizable definition of the system with a form for $G$
inspired on the LVS, where the wave function of the matter fields
is suppressed by some power of the volume. Thus, in this case,
splitting the $L$ sector by the moduli, ${\cal M}$, and the matter
fields, $Q$, we propose
\begin{equation}\label{GENFACT}
G(H,\bar H,L,\bar L)=G_H(H,\bar H)+G_{\cal M}({\cal M},\bar {\cal
M})+\epsilon G_{mix}(\phi,\bar \phi)\,,
\end{equation}
with $\phi^I=\lbrace H,{\cal M},Q\rbrace$. This is: the matter
fields $Q$ only enter in the suppressed part of $G$ which can
depend in all three kind of fields. This form implies that the
matter fields enter in the superpotential also in a suppressed
way, forbidding ${\cal O}(1)$ couplings, e.g., Yukawa like. For
the moment let us simply say that fields appearing on such a
coupling usually develop vanishing VEV's, which are not dangerous
for the factorizability and we postpone its study for the moment.
Let us see that a structure like the one defined in
eq.\eqref{GENFACT} indeed allows for a decoupling between the two
sectors by integrating out the $H$ superfields.\\
We use the superconformal approach
\cite{Kaku:1978nz,Kugo:1982mr,Kugo:1982cu,Ferrara:1983dh} to write
down the SUGRA Lagrangian since the tensor calculus is similar to
the one of rigid supersymmetry. The procedure introduces new
degrees of freedom collected in the conformal gravity multiplet
with components the graviton, the gravitino and two vector
auxiliary fields. Also a compensator chiral multiplet, $\Phi$, is
added. By fixing the scalar and spinor components of the
compensator and one vector auxiliary field one recovers the
symmetries of ordinary SUGRA. The Lagrangian, then, up to two
derivatives can be written as an integration over rigid
super-coordinates \cite{Kugo:1982mr,Kugo:1982cu,Ferrara:1983dh},
\begin{equation}\label{SUGRALwithGTheta}
{\cal L}=\int d\theta^4\left(-3 e^{-G/3}\Phi \bar \Phi\right)+\int
d\theta^2 \Phi^3+h.c.\,.
\end{equation}
The Berenzin integrals, however, are now deformed by extra-terms
with dependencies on the components of the gravity multiplet. We
will mainly be interested in the scalar component Lagrangian,
where these deformations are not relevant, so the calculation is
straightforward and easily written like in the global case. With
conventions for the metric signature $(+,-,-,-,-)$ and for the
chiral fields $\phi=(\phi,\psi,-F^\phi)$ the scalar Lagrangian
reads,\footnote{For simplicity of notation, here and throughout
the paper, we use the same notation for the chiral multiplets and
its lowest components, being clear from the context to which one
we are referring to.}
\begin{align}\label{SUGRAFULLLwithG}
{\cal L}=&G_{M\bar M}\partial_\mu \phi^M\partial^\mu\bar\phi^{\bar
M}+ G_M F^M \bar U+G_{\bar M}\overline F^{\bar M}U\cr
&+\left(G_{M\bar M}-\frac{1}{3}G_M G_{\bar M}\right)F^MF^{\bar
M}-3U\bar U-3 e^{\frac{G}{2}}(U+\bar U)\,,
\end{align}
where the gauge fixing $\Phi=e^{G/6}$, $F^\Phi=e^{G/6}U$ has been
implemented in order to get a canonical normalized
Einstein-Hilbert action \cite{Kugo:1982mr}, and we have adopted
the convention on the notation for the derivatives on the fields
$\partial_M G=G_M$, capital Latin letters running over all fields
$H$, ${\cal M}$ and $Q$. Contrary to the analysis performed in
\cite{Achucarro:2008sy,Gallego:2008qi} we will keep the auxiliary
fields in the game. This not only makes the SUSY nature of the
theory to be completely manifest but also will make the following
analysis neat, showing more powerful in the case with gauge
interactions. Then, to be consistent with SUSY we now integrate
out simultaneously both components of the $H$ multiplets, whose
e.o.m. are
\begin{align}
F^i:=~~& G_i \bar U+\left(G_{i\bar N}-\frac13G_i G_{\bar
N}\right)F^{\bar N}=0\,,\label{EOMFORF}\\
H^i:=~~&-G_{i\bar N}\partial^2\phi^{\bar N}+G_{iM\bar
N}\partial_\mu\phi^M\partial^\mu\phi^{\bar N}+G_{iM}F^M\bar
U+G_{i\bar M}F^{\bar M} U\cr &-\frac32 e^{G/2}(U+\bar
U)G_i+\left[G_{iM\bar N}-\frac13\big(G_{iM}G_{\bar N}+G_M G_{i\bar
N}\big)\right]F^MF^{\bar N}=0\,,
\end{align}
with convention of small Latin letters running only over the $H$
fields. The first equation together with the on-shell expression
for the compensator auxiliary field, $U=\frac13G_MF^M -e^{G/2}$,
leads to the on-shell expression for the auxiliary fields, that we
write as
\begin{equation}
G_i=e^{-G/2}G_{i\bar N} F^{\bar N}\,.
\end{equation}
Regarding the approximate factorizability of the function $G$ the
e.o.m. for the $H$ fields reads,
\begin{eqnarray}\label{EOMFACTORIZABLE}
G_{ij} F^j\bar U+G_{i\bar j}\overline F^{\bar j}U+G_{ij\bar k}F^j\overline F^{\bar k}-\frac13G_{ij}G_{\bar N}F^j\overline F^{\bar N}-\frac13G_MG_{i\bar j}F^M\overline F^{\bar j}&&\\
-\frac{3}{2}e^{G/2}(U+\bar U)G_i-G_{i\bar j}\partial^2 \bar
H^{\bar j}+G_{i\bar j\bar k}\partial^\mu \bar H^{\bar
j}\partial_\mu\bar H^{\bar k}&=&{\cal O}(\epsilon)\,,\nn
\end{eqnarray}
so that at leading order in $\epsilon$ the $Q$ fields completely
disappear meanwhile the ${\cal M}$ ones only appear in the
$G_MF^M$ terms. Here we see that although there is no mass
hierarchy between the two sectors the factorizable nature of the
$G$ function, and in particular of the K\"ahler potential, leads
to almost decoupled wave functions so that the rotation to normal
coordinates does not mix the two sectors. In other words, despite
the mass scales of both sectors are the same the kinetic energy
from the $L$ sector cannot, at first approximation, excite
fluctuations in the $H$ sector (see also \cite{Achucarro:2010jv}),
allowing for a two derivative approximation. Looking for slowly
varying solutions, using the on-shell expression for the $U$ field
and eq.\eqref{EOMFORF}, one gets
\begin{equation}\label{ApproxHEOM}
G_{ij} F^j\bar U+G_{ij\bar k}F^j\overline F^{\bar
k}-\frac12(U+3\bar U)G_{i\bar j}\overline F^{\bar
j}-\frac13G_{ij}G_{\bar N}F^j\overline F^{\bar
N}-\frac13G_MG_{i\bar j}F^M\overline F^{\bar j} ={\cal
O}(\epsilon)\,.
\end{equation}
These, have  solutions $F^i={\cal O}(\epsilon)$ implying, from
eq.\eqref{EOMFORF}, that the lowest component of $H$ be solution
for
\begin{equation}\label{HFflatness}
\partial_iG={\cal
O}(\epsilon)\,,
\end{equation}
so that once we require that $W_H$, from $G_H=K_H+\log|W_H|^2$, be
${\cal O}(1)$ necessarily implies
\begin{equation}\label{leadHFflatness}
\partial_iG_H=\frac{1}{W_H}(\partial_iW_H+\partial_iK_H W_H)={\cal
O}(\epsilon)\,,
\end{equation}
which are nothing but the leading $F$-flatness conditions one
finds by solving the e.o.m. obtained directly from the lowest
component potential \cite{Gallego:2008qi}. Of course, it might
happen that these solutions do not exist, or that these do not fix
all $H$ fields. We will suppose in the following that this is not
the case and all $H$ fields are fixed by \eqref{leadHFflatness}
avoiding issues like the one pointed out in \cite{Choi:2004sx}.
The solution for the lowest component of $H$, then, can be cast as
$H^i=H^i_o+\epsilon \Delta H^i$, where $H_o^i$ is the solution for
$\partial_i G_H=0$, which is $L$ independent.
\\
Plugging back these solutions in \eqref{SUGRAFULLLwithG} keeping
only up to ${\cal O}(\epsilon)$ factors, one gets, with the Greek
indices running only over $L$ sector fields,
\begin{eqnarray}\label{EFFLGENGFACT}
{\cal L}_{eff}&=&G_{\alpha\bar \beta}\partial_\mu
\phi^\alpha\partial_\mu\bar \phi^{\bar \alpha}+G_\alpha F^\alpha
\bar U+G_{\bar \alpha}\overline F^{\bar
\alpha}U\cr&&+\left(G_{\alpha\bar \beta}-\frac{1}{3}G_\alpha
G_{\bar \beta}\right)F^\alpha F^{\bar \beta}-3U\bar U-3
e^{\frac{G}{2}}(U+\bar U)+{\cal O}(\epsilon^2)\,,
\end{eqnarray}
where we have kept explicitly, in the last term, some ${\cal
O}(\epsilon^2)$ factors in order not to be forced to split the $G$
function and overload the notation. Notice that the only place
where the $H^i$ lowest component solution appear at ${\cal
O}(\epsilon^0)$ is in the last term. However, expanding around the
leading solution, $\partial_i G={\cal O}(\epsilon)$, the
corrections will appear only at the ${\cal O}(\epsilon^2)$, so we
can safely keep only the leading solution for the $H^i$, i.e.,
$H_o^i$, as well in the rest of the Lagrangian being at most
${\cal O}(\epsilon)$, and therefore any possible $L$ dependency of
the $H$ solution completely disappears.
\\
The effective theory is then described at leading order by the
Lagrangian explicitly written in eq.\eqref{EFFLGENGFACT}, that is
precisely the one of a SUGRA theory with K\"ahler invariant
function given by the original one but where the $H$ superfields
are frozen at their leading solutions,
\begin{equation}
G_{simp}({\cal M},\bar{\cal M},Q,\bar Q)=G(H_0,\bar H_0,{\cal
M},\bar {\cal M},Q,\bar Q)\,,
\end{equation}
with $H_0$ the constant chiral superfield with vanishing spinor
and auxiliary components, and scalar component the leading
solution of \eqref{HFflatness}. One can further check that the
SUSY breaking contributions from the $H$ sector are in fact
suppressed compared to the $L$ sector ones. Taking the canonical
normalized $F$-term, $F_c^M=|K_{M\bar M}\overline F^{\bar
M}F^{M}|^{1/2}$ no sum, we have
\begin{equation}
F^{\cal M}_c\sim{\cal O}(1)\,,~~F_c^Q\sim{\cal
O}(\epsilon^{1/2})\,,~~F^i_c\sim{\cal O}(\epsilon)\,.
\end{equation}
We are finding explicitly that the resulting theory obtained by
integrating out the $H$ fields, regardless the lack of a hierarchy
with the ${\cal M}$ fields, nor with the SUSY breaking scale
\cite{Brizi:2009nn},\footnote{Notice that the physical masses,
canonically normalized, and SUSY breaking scale, once the $L$
sector breaks it, are all of ${\cal O}(1)$.} can be described at
leading order in $\epsilon$ by a SUGRA theory whose Lagrangian
coincides with one of the simplified version. In particular, in
absence of matter fields the simplified version is valid at next
to leading order since for the moduli the Lagrangian is of ${\cal
O}(1)$. In this way we recover the results in
\cite{Achucarro:2008sy,Gallego:2008qi}, with the addendum of
making more manifest the SUSY nature of the resulting theory.
\\
We understand, moreover, that it is thanks to the weak coupling
between the two sectors that the solutions for the $F^i$ auxiliary
fields are suppressed, independently of the remaining dynamics,
and we can realize an effective SUSY theory although the masses
for the $H$ fields turn our to be of the same order of the SUSY
breaking scale.

\subsection{Large volume scenario}\label{ONLYCHIRALVSsect}

The ${\cal N}=1$ SUGRA theory obtained in type-IIB orientifold
compactifications, where the LVS are realized, is described by a
K\"ahler potential with general structure, including ${\cal
O}(\alpha^{\prime3})$ corrections \cite{Becker:2002nn}, given by
\begin{equation}\label{ModKPot}
K=-2\log\big({\cal V}+\xi\,(S+\overline S)^{3/2}\big)+{\cal
K}_{CS}\,,
\end{equation}
where ${\cal K}_{CS}$ depends only on the Dilaton, $S$, and
complex structure, $U^i$, but whose explicit form is irrelevant
for our study. The Calabi-Yau $3$-fold (CY) volume, ${\cal V}$, is
a K\"ahler moduli dependent function, such that in case of absence
of $\alpha^\prime$ corrections, encoded in the $\xi$ parameter,
the SUGRA theory has a no-scale nature.
\\
Although it has been studied some generic properties the theory
should satisfied in order to realize vacua with exponentially
large volumina \cite{Cicoli:2008va}, explicit realizations of
moduli stabilization and characterization of LVS properties are in
general based on the so called ``Swiss-cheese'' manifold
compactifications which probably encode all the main features of
the LVS models. In the simplest of such manifolds the compact CY
is realized as the hypersurface $\textbf{CP}^4_{[1,1,1,6,9 ]}$,
with $h^{2,1}=272$ complex structure moduli and $h^{1,1}=2$
K\"ahler moduli: one breathing mode, $T$, and a blow-up mode, $t$.
The volume is in this case \cite{Denef:2004dm},
\begin{equation}\label{CP4}
{\cal V}=\lambda\big(T_r^{3/2}-t_r^{3/2}\big)\,,
\end{equation}
where $\lambda=1/9\sqrt{2}$, but we leave it in this form in order
to have clearer formulae, and we have introduced the notation
$\phi_r=(\phi+\bar \phi)/2$ used here after. In the following we
stick to this form for the volume, mainly to match the existent
models but also to have a precise dependency on the $L$ fields
being crucial for the analysis. Our results, however, are
automatically extended to more complicated Swiss-cheese manifolds
since in general with more blow-up modes the volume is generalized
as ${\cal V}\sim T^{3/2}-\sum_i t_i^{3/2}$. At the qualitatively
level our results are also expected to hold for other kind of
manifolds.
\\
There is a flux induced tree level superportential given by
\cite{Gukov:1999ya}
\begin{equation}\label{GVWW}
W_{cs}=\int\Omega_3\wedge G_3\,,
\end{equation}
where $\Omega_3$ is the $(3,0)$-form of the compact manifold, and
$G_3=F_3+iS\, H_3$ the combined three-form-flux. The K\"ahler
moduli, instead, appear in the superpotential only through
non-perturbative dynamics, which can also depend on the other
moduli. In the spirit of a Two-Step moduli stabilization the
Dilaton and Complex structure moduli are regarded as frozen due
the superpotential \eqref{GVWW} with larger dynamics compared to
the one encoded in the non-perturbative part. It was found in
\cite{Gallego:2008qi,Gallego:2009px} that in order this procedure
be reliable for generic K\"ahler potential one needs to require a
tuning at the vacuum for the tree level superpotential. Indeed,
this tuning implies a mass hierarchy which warranties the
decoupling between the two sectors. In a more natural scenario the
classical superpotential is to get ${\cal O}(1)$ VEV, and the
decoupling relies in the factorizability of the $G$ function
\cite{Achucarro:2008sy,Gallego:2008qi}. On the side of the
superpotential the mixing between the complex and the K\"ahler
moduli appears only by non-perturbative dynamics, so are naturally
suppressed and the approximate factorizability, as defined in
eq.\eqref{NEARFACTG}, is safe. On the other hand, the
$\alpha^\prime$ corrections of the K\"ahler potential, breaking
the factorizability, are not naturally suppressed. However,
whenever we stand at points of the moduli space such that the
compact manifold volume is quite large, like in the LVS, these
corrections are small an approximate factorizable system is
realized. Notice that contrary to the KKLT like models studied in
\cite{Gallego:2008qi,Gallego:2009px}, besides the suppression
factor coming from the non perturbative dynamics, there is a
second one dictated by the size of the volume. This two {\it a
priori} unrelated factors are however connected through the
stabilization of the moduli and this relation depends on the
particular model, making not viable a generic study as was
possible for the KKLT like family of models.
\\
As said before the presence of matter fields breaks the
factorizability in the K\"ahler potential, but in the extended
sense given in eq.\eqref{GENFACT} one can still get near
factorizability between the $H$ sector, Complex structure and
Dilaton moduli, and the $L$ sector, K\"ahler moduli and charged
fields. Indeed the K\"ahler potential for the matter fields reads
\cite{Conlon:2006tj}
\begin{equation}\label{KahlerQ}
 K\supset \frac{Z}{{\cal V}^n}|Q|^2\,,
\end{equation}
where $Q$ is a generic charged field, omitting for simplicity
possible index contractions, $Z$ is a real function of the Complex
structure and the blow-up K\"ahler moduli $t$, and the modular
weight $n$ is a positive number. Depending on the cycle the
$D7$-branes wrap, where the matter is realized, the moduli
dependent function $Z$ trivializes or not, and the suppression
factor is a power of the volume or just a power of the breathing
mode $T$. For our purposes this schematic form is enough, noticing
again that around points in the moduli space where the size of the
volume is quite large the appearance of the $Q$ in the generalized
function $G$ is suppressed as far its dependencies in the
superpotential be also suppressed. We have, then, in general a
third suppressing factor depending on the modular weight.
\\
Let us start by considering the simplest case where $W$ is $Q$
independent, so that the superpotential takes the form, with $A$ a
function in general of the $H$ fields,
\begin{equation}
W=W_{sc}+A\,e^{-a\,t}\,,
\end{equation}
with the non-perturbative part of the superpotential depending
only on the {\sl a priori} regarded small modulus. More precisely
the breathing mode is supposed extremely large so that such
dependencies vanish. If we now suppose that the K\"ahler moduli
stabilization follows like in the pure moduli case, as actually
does and even it turns out to happen in realistic models seed by
this one (see appendix \ref{appLVSwithMatter}), the non
perturbative suppression factor coincides with the inverse of the
volume and we can define a single suppressing factor given by
$\epsilon\sim 1/{\cal V}\sim 1/T^{3/2}\sim A\,e^{-a\,t}$,
meanwhile $t={\cal O}(1)$. The analog of eq.\eqref{ApproxHEOM}
then reads
\begin{align}\label{LargEOM}
G_{ij}F^i\bar U+G_{ij\bar k}F^iF^{\bar k}-\frac13 G_{ij}G_{\bar
N}F^jF^{\bar N}-\frac13G_{i\bar j}G_M F^{\bar j}F^M -\frac12
(U+3\bar U) G_{i\bar j}F^{\bar j}\cr \sim ( U+F^{ j}+G_{ N}F^{
N})G_{i\alpha} F^\alpha
+G_{i\alpha\bar\beta}F^{\alpha}F^{\bar\beta}\,,
\end{align}
where as before the capital letters, $M$ and $N$ run over all
fields, $i$, $j$, over the $H$ ones, and the Greek indexes over
the K\"ahler moduli and the $Q$ fields. In the r.h.s. we have been
sloppy with the holomorphic indexes and their contraction being
only interested in the leading scaling in $\epsilon$: for example
for $t$, in case $Q$ has a vanishing VEV, we have $G_{t\bar H}\sim
\epsilon^2$ meanwhile $G_{tH}\sim \epsilon$, so we take the last
one for the analysis. The vector and matrix shown in the r.h.s. of
eq.\eqref{LargEOM} have the scalings on $\epsilon$,
\begin{align}
G_{i\alpha}\sim
&\Big(\epsilon^{2/3}(\epsilon+\epsilon^{n}Q^2),\epsilon+\epsilon^{n}Q^2,
\epsilon^{n} \bar Q\Big)\,,\cr G_{i\alpha\bar \beta}\sim &
\begin{pmatrix}
 \epsilon^{4/3}(\epsilon+\epsilon^{n}Q^2) & \epsilon^{2/3}(\epsilon^2+\epsilon^{n}Q^2)&\epsilon^{2/3+n} Q  \\
  \epsilon^{2/3}(\epsilon^2+\epsilon^{n}Q^2) &  \epsilon+\epsilon^{n}|Q|^2 & \epsilon^{n}Q \\
  \epsilon^{2/3+n} Q   & \epsilon^{n}Q & \epsilon^{n}
\end{pmatrix}\,,
\end{align}
$i$ fixed and choosing the ordering $\lbrace T,t,Q\rbrace$. Having
in mind that the on-shell solution for the auxiliary fields in the
$L$ sector are necessary suppressed by powers of $\epsilon$ and
looking for approximate SUSY solution in the $H$ sector the
leading solution for $F^i$ auxiliary fields has the schematic form
\begin{align}\label{FiNoGauge}
F^i\sim G_{i\alpha} F^\alpha
\sim\epsilon^{2/3}(\epsilon+\epsilon^{n}Q^2)
F^T+\big(\epsilon+\epsilon^{n}Q^2\big) F^t+\epsilon^n QF^Q\,,
\end{align}
being again sloppy with the holomorphicity of the indexes, in fact
the solution will mix them up in general. This leads us again to
conclude that the leading solution for the $H$ lowest component is
dictated by the the $F^i$-flatness conditions,
\begin{equation}
e^{G/2}G_i=G_{i\bar N}F^{\bar N}\sim
\epsilon^{2/3}(\epsilon+\epsilon^{n}Q^2)
F^T+\big(\epsilon+\epsilon^{n}Q^2\big) F^t+\epsilon^n QF^Q\,,
\end{equation}
then, although $e^{G/2}\sim\epsilon$, the on-shell expressions for
the $F^\alpha$ show that $G_i$ vanishes at leading order in
$\epsilon$. Indeed equation \eqref{FiNoGauge} tells us about the
suppression of the SUSY breaking contribution from the $H$ sector.
As before, we consider that $G_H={\cal K}_{cs}+\log|W_{sc}|^2$ is
such that allows for such solutions and fixes the whole set of $H$
fields. Plugging back these in the Lagrangian one has
\begin{eqnarray}\label{effL}
{\cal L}_{eff}&=&G_{\alpha\bar \beta}\partial_\mu
\phi^\alpha\partial_\mu\bar \phi^{\bar \alpha}+G_\alpha F^\alpha
\bar U+G_{\bar \alpha}\overline F^{\bar
\alpha}U\cr&&+\left(G_{\alpha\bar \beta}-\frac{1}{3}G_\alpha
G_{\bar \beta}\right)F^\alpha F^{\bar \beta}-3U\bar U-3
e^{\frac{G}{2}}(U+\bar U)+{\cal O}\big((F^i)^2\big)\,,
\end{eqnarray}
where as before we can keep only the leading solution for the
scalar component of $H$, i.e., $\partial_i G_H=0$, which are
independent of the $L$ fields, being the further corrections
negligible. As pointed out before the corrections ${\cal
O}\big((F^i)^2\big)$ will have not only the SUSY combinations
$F\bar F$ but also some pure holomorphic and pure non-holomorphic
terms that reveal the non-SUSY nature of this effective
Lagrangian. We should compare all the ${\cal O}\big((F^i)^2\big)$
corrections with terms coming from a SUSY Lagrangian constructed
from $G_{sim}(T,\bar T,t,\bar t,Q,\bar Q)=G(H_0,\bar H_0,T,\bar
T,t,\bar t,,Q,\bar Q)$, with $H_o$ a superfield with vanishing
spinor and auxiliary field components and whose lowest component
is fixed by the leading solution to the $F^i$-flatness conditions,
which are again $L$ independent. This Lagrangian is precisely the
one explicitly written in \eqref{effL} with some subleading
corrections due to the fact that the correct solution for the
scalar component in shifted from the leading SUSY value.
\\
Let us perform explicitly the analysis of the term $|F^t|^2$, so
to make the procedure clear, and furthermore spot possible
numerical factors necessary to precisely understand the numerical
tests performed later on in the appendices. Taking the term
$|F^t|^2$ we have that in the simplified version it is lead by
\begin{equation}
G_{t\bar t}\sim\frac{3\lambda}{8\sqrt{t_r}{\cal
V}}+\frac{\partial_t\partial_{\bar t}Z}{{\cal
V}^n}|Q|^2\sim\epsilon+\epsilon^n |Q|^2\,.
\end{equation}
This term is also generated in the effective Lagrangian by
$|G_{it}F^t|^2$, with leading terms $G_{it}\sim a\,
e^{-a\,t}\partial_i A+\frac{\partial_t\partial_{i}Z}{{\cal
V}^n}|Q|^2\sim \epsilon+\epsilon^n |Q|^2$, so there is a further
contribution not present in the naive Lagrangian which goes like
$(\epsilon+\epsilon^n |Q|^2)^2|F^t|^2$, but is suppressed by
$\Delta= {\cal O}(\epsilon,\epsilon^n |Q|^2)$. In order to keep
track of the numerical factors one has to recall that
$A\,e^{-a\,t}=\tilde A \lambda \sqrt{t}\frac{\lambda |W_o|}{a}
\epsilon$ (see appendix \ref{appLVSwithMatter}), with $\tilde A$
encoding possible extra factors missed by the analytic procedure,
so that in the $Q$ independent parts of the corrections one finds
an enhancement. Being precise one finds $\Delta\sim \frac83
\lambda t^{3/2}\tilde A^2\epsilon$. The same analysis is done for
all terms, which in the simplified Lagrangian are ruled by the
derivatives
\begin{align}\label{GderivscalingLSV}
G_\alpha\sim
&\Big(\epsilon^{2/3}(1+\epsilon^{n}|Q|^2),\epsilon+\epsilon^{n}|Q|^2,
\epsilon^{n} \bar Q\Big)\,,\cr G_{\alpha \bar\beta}\sim
&\begin{pmatrix}
 \epsilon^{4/3}(1+\epsilon^{n}|Q|^2) & \epsilon^{2/3}(\epsilon+\epsilon^{n}|Q|^2)&\epsilon^{2/3+n} Q  \\
  \epsilon^{2/3}(\epsilon+\epsilon^{n}|Q|^2) &  \epsilon+\epsilon^{n}|Q|^2 & \epsilon^{n}Q \\
  \epsilon^{2/3+n} \bar Q   & \epsilon^{n}\bar Q & \epsilon^{n}
\end{pmatrix}\,.
\end{align}
In order to compare terms that are not originally present in the
simplified version, e.g., the pure holomorphic $(F^T)^2$ term, we
replace one of the auxiliary field to its scaling around the
vacuum, which from eq.\eqref{GderivscalingLSV} with $F^\alpha\sim
e^{G/2}G_{\alpha\bar\alpha}^{-1}G_{\bar \alpha}$ no sum and
\eqref{FiNoGauge}, are
\begin{equation}\label{auxLVS}
F^T\sim \epsilon^{1/3}\,,~~F^t\sim \epsilon\,,~~F^Q\sim \epsilon
Q\,,\Longrightarrow F^i\sim\epsilon^2+\epsilon^{n+1} Q^2\,,
\end{equation}
so that
\begin{equation}
(F^i)^2\supset\big( \epsilon^2+\epsilon^{n+1}
Q^2\big)\left(\epsilon^{2/3}(\epsilon+\epsilon^{n}Q^2)
F^T+\big(\epsilon+\epsilon^{n}Q^2\big) F^t+\epsilon^n
QF^Q\right)\,,
\end{equation}
and compare the results with the corresponding one, e.g., $F^T$,
regarding the scaling for the compensator auxiliary field as
$U\sim \epsilon$, resulting from its on-shell expression,
$U=\frac13 G_{\bar M}F^{\bar M}-e^{G/2}$. In this way one finds
that all corrections encoded in the ${\cal O}\big((F^i)^2\big)$
terms are suppressed by ${\cal O}(\epsilon, \epsilon^n Q^2)$. As
explained in appendix \eqref{appNoQinW} in order the stabilization
of the moduli indeed proceeds as in the standard case and we can
relay on the given scalings, one should impose the condition
$\langle Q^2\rangle\lesssim \epsilon^{1-n}$, so the largest
corrections are of ${\cal O}(\epsilon)$, independent of the
modular weight $n$!. In this way we generalize the results
obtained in \cite{Gallego:2008qi} working now in presence of
matter fields and keeping track of the auxiliary fields.
\\
To close this section let us give a look at the canonical
normalized auxiliary fields VEV, showing that indeed the SUSY
breaking contribution from the $H$ sector is suppressed,
\begin{equation}
F^T_c\sim
\epsilon\,,~~F^t_c\sim\epsilon^{3/2}\,,~~F^Q_c\sim\epsilon^{1+n/2}Q\lesssim
\epsilon^{3/2}\,,~~F^i_c\sim\epsilon^2\,,
\end{equation}
as obtained from eqs.\eqref{FiNoGauge}, \eqref{GderivscalingLSV}
and \eqref{auxLVS}.

\section{Chiral and vector multiplets}\label{VectorAndChiralSect}

We now turn on gauge interactions, that can be seen as some gauged
isometries of the scalar manifold, which for a ${\cal N}=1$ SUGRA
is of the K\"ahler type. The chiral multiplet transformations are
then defined by $\delta \phi^M = \Lambda^A X^M_A$, $\delta
\bar\phi^{\bar M} =\overline\Lambda^A \overline X^{\bar M}_A$,
with $\Lambda^A$ the gauge chiral parameter, $X_A^M$ ($\overline
X^{\bar M}_A$), the holomorphic (antiholomorphic) Killing vectors
generating the (gauged) isometry group ${\cal G}$, and the gauge
group indices running over $A=1,\,2\,\ldots {\rm adj}({\cal G})$.
The associated gauge vector fields transform as $\delta
V^A=-i(\Lambda^A-\overline\Lambda^A)$, inducing a total
transformation on $G$ of the form $\delta
G=\Lambda^AX_A^IG_I+\overline\Lambda^A\overline X^{\bar
I}_AG_{\bar I}-i(\Lambda^A-\overline\Lambda^A)G_A=0$, with $G_A$
denoting derivatives of $G$ with respect to the vector multiplets,
and as before the Latin letters running over the chiral
multiplets. Gauge invariance of the system then reads\footnote{We
have not included constant Fayet-Iliopoulos terms $\xi_A$ in
eq.\eqref{GgaugeInv} since they seem not to appear in theories
raised from string compactifications.}
\begin{equation}\label{GgaugeInv}
G_A=-iX^I_AG_I\,,
\end{equation}
a relation that we will use in the following. Notice that if we
want to fix the $H$ fields by the conditions $G_i=0$, all $H$
fields should be necessarily neutral otherwise by gauge
invariance, eq.\eqref{GgaugeInv}, the equations turn out to be
linearly dependent and leave some unfixed
directions.\footnote{Notice that one might generalize the way the
$H$ fields are fixed by regarding two gauge sectors, say ${\cal
G}={\cal G}_{\tilde A}\otimes{\cal G}_{ A}$, so that the $H$ are
charged under ${\cal G}_{\tilde A}$ but neutral under ${\cal
G}_A$, and the opposite for the $L$ fields. Then fix the $H$
fields by the requirement $G_i=0$ and $G_{\tilde A}=0$. This is,
besides the $F$-flatness the $D$-flatness conditions. In this case
some of the $H$ field are then stabilized by gauge dynamics and
not by the superpotential, as usually regarded for the $H$ sector
in string compactification, so we left this situation out of our
analysis.} This requirement also avoids the possibility the $H$
fields be sourced by the gauge fields in the effective theory. We
set, then, in the following $X^i_A=0$.
\\
Using the conformal formalism the Lagrangian of
eq.\eqref{SUGRALwithGTheta} is first of all corrected by the $G$
dependency on the vector multiplets $V^A$, promoting to covariant
derivatives the derivatives in the kinetic terms, i.e.,
\begin{equation}
\partial_\mu \phi^I\to \partial_\mu\phi^I+X^I_AV^A_\mu\,.
\end{equation}
Further terms for \eqref{SUGRALwithGTheta} come from the chiral
Lagrangian and from the kinetic part of the vector multiplets
Lagrangian. Working in the Wess-Zumino gauge, denoting the vector
auxiliary fields by $D^A$ one has the following extra terms to the
ones in eq.\eqref{SUGRAFULLLwithG} \cite{Kaku:1978nz}
\begin{equation}\label{DtermL}
{\cal L}\supset G_AD^A+\frac12 h_{AB}D^AD^B\,,
\end{equation}
with $h_{AB}=Re(f_{AB})$, $f_{AB}$ the field dependent gauge
kinetic functions.

\subsection{Nearly factorizable models}

As before let us first tackle the case of the generic nearly
factorizable models where the situation follows as before with the
e.o.m. for $F^i$ auxiliary fields not changed, eq.\eqref{EOMFORF},
and for the scalar components the analogous of
eq.\eqref{EOMFACTORIZABLE} now reads,
\begin{align}\label{ApproxHEOMGau}
-G_{i\bar j}\partial^2 \overline H^{\bar j}+G_{ij\bar
k}\partial_\mu H^j\partial^\mu\overline H^{\bar k}+G_{ij} F^j\bar
U+G_{ij\bar k}F^j\overline F^{\bar k}-\frac12(U+3\bar U)G_{i\bar
j}\overline F^{\bar j}&\cr-\frac13G_{ij}G_{\bar N}F^j\overline
F^{\bar N}-\frac13G_MG_{i\bar j}F^M\overline F^{\bar
j}+\frac12D^AD^B\partial_i h_{AB} &={\cal O}(\epsilon)\,,
\end{align}
obtained by using the relation \eqref{GgaugeInv}, the neutrality
of the $H$ fields, so that $\partial_i G_A=-iX_A^\alpha G_{\alpha
i}$, and regarding that none of the Killing vectors $X^\alpha$ get
anomalous large values, i.e., ${\cal O}(X^\alpha_A)\sim 1$. We
have, then, that the presence of the last terms makes the solution
$F^i\sim \epsilon$ no longer automatic; indeed, the e.o.m. tells
us that whenever the $D$-term gets a non-vanishing VEV, which is
naturally of ${\cal O}(1)$, it back reacts on the $H$ sector, even
if neutral, inducing a large $F^i$-term. In particular the $F^i$
auxiliary fields solutions would depend strongly on the $L$
fields, therefore cannot be simply neglected, as a Two-Step
stabilization requires, and have to be properly integrated out.
Notice, moreover, that the mixing between the $F$ and the $D$
auxiliary fields makes again explicit the non-SUSY nature of the
effective theory, which on the scalar side still is a two
derivative description.\footnote{A similar analysis was done in
\cite{Brizi:2009nn} studying the SUSY description of effective
theories.}
\\
The decoupling would be still safe if somehow the dependency of
the gauge kinetic function on the $H$ superfields is absent or
suppressed by say $\epsilon$. In this case things turn out to be
like in the case without gauge dynamics as the solution $F^i\sim
\epsilon$ is again valid, and the leading solution for the lowest
component is again dictated by the leading $F$-flatness
conditions, that do not depend on the $L$ sector. Thus, the
simplified theory, where the $H$ superfields are regarded as
frozen at their leading solution, is reliable at leading order, as
far as the $H$ superfields are neutral and the frozen value is, up
to ${\cal O}(\epsilon)$ corrections, the leading solution to the
$F^i$-flatness
condition.\\
Notice that we have not distinguished between the cases with and
without breaking of the gauge symmetries. In the study done in
\cite{Gallego:2009px} it was important to keep track of the gauge
symmetry breaking as some of the fields in the $L$ sector,
regarded as light, acquire heavy masses via the $D$-term dynamics.
Then, one needs to distinguish these fields from the low energy
spectrum when speaking about an effective theory integrating these
fields out, not being in general possible to freeze them out,
together with the $H$ fields. In our case however, modes acquiring
masses through this mechanism are not heavier than other fields in
the $L$ sector. Indeed, regarding the eigenvalues of $h_{AB}$ of
${\cal O}(1)$, their masses should be of the order of the
eigenvalues of vector field mass matrix associated to the broken
generators,
\begin{equation}
M^2_{AB}=2G_{I\bar I} X^I_A\overline X^{\bar I}_B\,,
\end{equation}
which are clearly of ${\cal O}(1)$. Therefore, there is not need
of spotting such fields and no formal distinction appears if the
symmetry is broken or not. There might appear a small hierarchy
with such fields, like in the case $N_f<N_C$ worked out in section
\ref{appNewModel}. In this case one can proceed like done in
\cite{Gallego:2009px} by integrating the massive vector
superfields through the superfield vector equation $\partial_VK=0$
\cite{Brizi:2009nn} and choosing a convenient gauge fixing, fixing
the value of one of the charged fields with a non-vanishing
component in the would-be Goldstone direction. Since the K\"ahler
potential has a factorizable form and the $H$ multiplets are
neutral the dependency of the solution on the $H$ fields will be
necessarily suppressed. Plugging back the solution into the
original K\"ahler potential leads then to a new K\"ahler potential
which however has still a factorizable form, and the same analysis
can be performed with only the unbroken symmetries. However, since
in our models is the weak coupling between the sectors and not the
mass hierarchy what plays the important role in the decoupling,
these issues are not relevant for our study, being this more clear
if we remember that the effective theory we are speaking about is
not a proper one in the Wilson sense.
\\
We find, then, that in order Two-Step procedure  be reliable in
general\footnote{We will show in later sections that what really
matters is the VEV of the $D^A$, so decoupling is still save
whenever this VEV is tiny.} for this generic factorizable models,
a further restriction is required, namely a suppressed dependency
of the gauge kinetic function on the fields to be frozen,
something that cannot be justified in general for realistic
situations. Nevertheless, it is clear that this general setup as
stands so far cannot be connected to a real situation as SUSY is
broken at Planck scale. Interestingly enough, in the next section
we find that in realistic realizations of nearly factorizable
models, namely the LVS, this restriction is avoided naturally.

\subsection{Large volume scenarios}

In type-IIB orientifold compactifications with $D7$-branes the
role of the gauge kinetic functions is played by the K\"ahler
moduli parameterizing the size of the cycles where the branes
wrap. The presence of fluxes, however, induces corrections on the
gauge kinetic functions that depend on the Dilaton modulus
\cite{Lust:2004fi}, which are supposed to be frozen in the
Two-step stabilization spirit. Thus, the condition we found for a
reliable decoupling in generic nearly factorizable models is not
naturally realized. Since the mixing between the $F^i$ and the
$D^A$ auxiliary fields, which is what rises this problem, still
appears, if decoupling indeed applies its justification should
come from other means. In the previous case, however, we were
forced to strengthen the requirements on the models from the fact
that the $D^A$ are naturally of ${\cal O}(1)$, as seen by their
on-shell expression,
\begin{equation}\label{Donshell}
D^A=-h^{AB}G_A=ih^{AB}X^\alpha_AG_\alpha\,,
\end{equation}
with the matrix $h^{AB}\equiv h_{AB}^{-1}$, since apart from the
matter field contribution, of ${\cal O}(\epsilon)$, $D^A\sim
G_\alpha={\cal O}(1)$. Even more, this scaling naturally holds at
the vacuum, where the following dynamical relation is satisfied
\cite{GomezReino:2007qi},
\begin{equation}
\frac12\left(M_{AB}^2+h_{AB}\big(G_{I\bar J}F^IF^{\bar J}-
e^G\big)\right)D^B=G_{AI\bar J}F^IF^{\bar J}\,,
\end{equation}
everything being naturally of ${\cal O}(1)$. On the other hand,
for our type-IIB setup all $G_\alpha$ are suppressed by powers of
the volume, so in a LVS things seems to be controlled on favor of
the decoupling, but even though the decoupling is still not
trivially clear. Moreover like in the pure chiral fields case a
complete generic analysis is not possible, so we will follow the
study separately for explicit models with matter fields realizing
moduli stabilization in LVS with vanishing cosmological constant
\cite{Cremades:2007ig,Krippendorf:2009zza}. In this models the
main effects from gauge dynamics, besides the non-perturbative
superpotential, come from Abelian sectors under which the moduli
get charged, so in the following we only consider tree level
effects coming from such $U(1)_X$ group. This do not loose the
generality of the study as the analysis of other sectors follows
in the same way and even more straightforward as the moduli are
neutral under those sectors and the dynamics are simpler. As a
matter of fact it is common to neglect such non-Abelian sectors by
working in a meson field description following $D$-flat
directions.
\\
As usual the symmetry is linearly realized for the matter like
fields, instead the moduli, controlling the gauge kinetic
functions, can develop a non-linear realization due to a
Green-Schwarz (GS) mechanism that cancels pseudoanomalies of the
Abelian sectors. In the case of type-IIB superstring with
$D7$-branes the K\"ahler moduli get charged once world volume
magnetic fluxes are turned on.\footnote{To be precise a
topological condition should be realized, namely that the
$4$-cycle whose volume is characterized by the K\"ahler modulus
has an intersection with the $2$-cycle where the world volume flux
is non-trivial, in order such a charge be not null
\cite{Haack:2006cy,Jockers:2005zy}.} The $U(1)_X$ symmetries in
the chiral superfields are then described by the holomorphic
Killing vectors,
\begin{equation}
X^I_X=i q_X^I \phi^I~\mbox{no sum}\,,~~X^{T^j}_X=i\frac12
\delta_X^{T^j}\,,
\end{equation}
with $q_X^I$ the charge of the chiral superfield $\phi^I$ and the
GS coefficient $\delta^{T^j}_X$ a real parameter associated to the
modulus $T^j$. The gauge invariant K\"ahler potential for the
moduli has then the functional dependency
\begin{equation}
K_{mod}=K\Big[T^j+\overline T^j+\frac12\delta_X^{T^j} V^X\Big]\,,
\end{equation}
with $V^X$ the vector superfield associated with the $U(1)_X$
symmetry. To make simpler the analysis we will regard that only
one of the K\"ahler modulus gets charged, but in general is
possible that several of them develop non trivial charges. The
main effect for moduli stabilization issues is the new
contribution induced in the $D$-term dynamics which enters like a
Fayet-Iliopoulos (FI) term but now field dependent
\cite{Cremades:2007ig,Krippendorf:2009zza,Binetruy:1996uv,ArkaniHamed:1998nu,Dudas:2007nz,Gallego:2008sv,Dudas:2008qf}.

\subsubsection{Charged large modulus}\label{TChargedSect}

The simplest implementations of this kind of models were developed
in \cite{Cremades:2007ig} with the large modulus
charged.\footnote{Although the authors of \cite{Cremades:2007ig}
also study models with a charged small modulus, these turn out to
lack of realizing a vanishing cosmological constant.} Although the
explicit realization of the models might have several matter
fields for our purposes their final out-shot can be cast using
only one $Q$, so that
\begin{equation}
X_X^T=i\frac12\delta^T\,,~~X^t_X=0\,,~~X_X^Q=q_Q Q\,.
\end{equation}
By gauge invariance the superpotential does not depend on the $Q$
field so the $F$-term part of the scalar Lagrangian goes like in
the case without gauge interactions, and it is possible to
stabilize the K\"ahler moduli like in the pure moduli case. The
scalings shown in eq.\eqref{GderivscalingLSV} and
eq.\eqref{auxLVS} then hold, and new scalings needed for the
analysis are, using eq.\eqref{GgaugeInv},
\begin{equation}\label{Gxscalings}
G_X\sim h_X\,D^X\sim
 \delta_T\epsilon^{2/3}+\epsilon^n|Q|^2\,,~~G_{Xi}=-i X_X^\alpha K_{\alpha i}\sim \delta_T\epsilon^{5/3}+\epsilon^{n} |Q|^2\,.
\end{equation}
Thus a small size for $D^X$ is implied and again is possible to
find a suppressed solution for the $F^i$ auxiliary fields, using
the on-shell expression for the $D^X$ auxiliary fields,
\begin{align}\label{FiTcharged} F^i&\sim G_\alpha F^\alpha+ 
\frac{1}{\epsilon}G_{X}D^X\cr
&=\epsilon^{2/3}(\epsilon+\epsilon^{n}Q^2)
F^T+\big(\epsilon+\epsilon^{n}Q^2\big) F^t+\epsilon^n
QF^Q+(\delta_T\epsilon^{-1/3}+\epsilon^{n-1} |Q|^2)D^X\,,
\end{align}
with the last term $1/\epsilon$ factor coming by regarding the
compensator scaling $U\sim \epsilon$. This last term apparently
couple strongly the two sector, and might give rise to $L$
dependencies in the $H$ solution; however, it turns out to be
inoffensive as explained bellow, so the effective Lagrangian is
again given by the simplified version plus corrections of the form
\begin{equation}\label{F2correctgauge}
{\cal O}(F^i)^2\sim (F^i_F)^2+\frac{1}{\epsilon}G_X D^X
F^I_F+\frac{1}{\epsilon^2}G_X^2(D^X)^2\,,
\end{equation}
where by $F_F^i$ we denoted the pure $F$ auxiliary fields
contribution, eq.\eqref{FiNoGauge}.
\\
The analysis of the corrections to the simplified version from the
$(F^i_F)^2$ part is exactly the same as the one with no gauge
interaction since the $F$-term part of the Lagrangian does not
change. Remains, then, only to check the new features coming from
the last two terms which enter as corrections to the $F^\alpha$,
$D^X$ and $(D^X)^2$ terms.
\\
Under this setup there are two possible scenarios, depending on
the sign of the $Q$ charge, differing by the VEV of $Q$ which is
ruled mainly by the minimization of the $D$-term dynamics,
eq.\eqref{DtermL}, which once the $D$ auxiliary fields are
integrated out, eq.\eqref{Donshell}, reads
\begin{equation}\label{VDpot}
V_D=\frac{1}{2\, h_X}(G_X)^2=\frac{1}{2\,h_X}\left(
\frac12\delta^T
\partial_TK+q_Q \frac{Z}{{\cal V}^n}|Q|^2 \right)^2\,,
\end{equation}
where the gauge invariance of the superpotential,
$X^I\partial_IW=0$ is used.
\\
The leading behaviour of the first term in \eqref{VDpot} is given
by $\partial_T K\sim -1/T_r<0$, and if the charge is of the
opposite sign of the GS coefficient the minimization of $V_D$
leads to a vanishing VEV for $Q$. Then, at the minimum the
$D$-term potential scales like $V_D\sim
\frac{1}{h_X}\left(\frac{\delta^T}{T_r^2}\right)^2$, and so in
order to uplift the $V_F=-{\cal O}(\epsilon^3)$ vacuum without
sweeping away the moduli stabilization, one requires a tuning
$\delta^T\sim \epsilon^{5/6} \sqrt{h_X}$. This, together with the
vanishing of $\langle Q\rangle$, implies that the dangerous
coupling in \eqref{FiTcharged} is in fact suppressed even in the
extreme case when $h_X\sim T_r\sim \epsilon^{-2/3}$. From the
effective Lagrangian point of view we have that since
$G_XD^X/\epsilon\sim \epsilon^2$, independently of the scaling of
$h_X$, and $F^i_F\sim \epsilon^2$ as before, then
$G_XD^XF^i_F/\epsilon$ leads to correction to the pure $F$-term
part similar to the ones coming from $(F^i_F)^2$ and which we
learned before are suppressed by ${\cal O}(\epsilon)$. These also
rise corrections to the $G_X D^X$ term of the form $\epsilon
G_XD^X$. The term $h_X (D^X)^2$ is corrected by the last term in
\eqref{F2correctgauge} that using \eqref{Gxscalings} and the
constrain on $\delta^T$ is of the form $\epsilon h_X (D^X)^2$. We
find, then, that all new possible corrections continue to be
suppressed at least by ${\cal O}(\epsilon)$.
\\
In case the charge turn out to be of the same sign of the GS
coefficient the minimization of the $D$-term potential tends to a
cancellation lead by a non-vanishing VEV for $Q$, which now scales
like $\langle Q^2\rangle\sim \epsilon^{2/3-n}\delta^T$. The
uplifting now proceeds from the $F$-term part with a term scaling
like $V_F\supset \epsilon^{2+n} Q^2$, so in  order to uplift
without spoiling the moduli stabilization a tuning $\delta^T\sim
\epsilon^{1/3}$ is required. Due to the $Q$ dependency in the
$F$-term potential the cancellation of the $D$-term part is not
exact, finding in fact that at the vacuum $D^X\sim \epsilon ^2$,
independently of the scaling of $h_X$, and therefore $G_X\sim h_X
\epsilon^2$. We have then that $G_X D^X\sim h_X \epsilon^4$ and
even in the worst case $h_X\sim T\sim\epsilon ^{-2/3}$ it is
smaller that in the previous case so the correction to the pure
$F$-term part are further suppressed, more precisely are
suppressed at least by ${\cal O}(\epsilon^{4/3})$, the same
happening to the corrections to the $h_X(D^X)^2$ terms. Since the
change in $F^i_F$ is negligible, $F^i_F\sim \epsilon^2+ h_X
\epsilon^3\sim \epsilon^2$, the corrections to $G_XD^X$ are still
of ${\cal O}(\epsilon)$.\footnote{\label{importantFN}One might
wonder if it is correct to take the scaling of $D^X$ at the
vacuum, which for this last case is drastically smaller than the
off-shell one. In fact, taking the off-shell scaling
\eqref{Gxscalings} with the constrain on $\delta^T$ one would find
corrections of ${\cal O}(\epsilon^{2/3})/h_X$ so that in the
situation $h_X\sim t\sim 1$ the corrections would be ${\cal
O}(\epsilon^{2/3})$ not of ${\cal O}(\epsilon)$ as claimed above.
We have check numerically that indeed this is not the case by
taking a sample with $h_X=t_r$, finding still corrections of
${\cal O}(\epsilon)$.} Interestingly we still find that the
corrections are independent of the modular weight and
the decoupling is in general safe. 

\subsubsection{Charged small modulus}

Once one allows a charged small modulus things turn out to be more
reach and intricate as gauge invariance imply the appearance of
matter like fields in the superpotential. Thus the stabilization
of the moduli should be revisted, possibly affecting the
scalings used above. 
\\
Again the minimization of the $D$-term potential, which now takes
the form
\begin{equation}
V_D\sim\frac{1}{h_X}\left(-\frac{\delta^t}{2}\partial_tK+q_Q
Q\partial _Q K \right)\sim
\frac{1}{h_X}\left(-\delta^t\epsilon+q_Q\epsilon^n |Q|^2\right),
\end{equation}
plays a crucial role in the stabilization of the matter. Notice,
however, the change in the sign for the field dependent FI term
which makes the implementation of such setup quite different to
the usual SUGRA FI model
\cite{Binetruy:1996uv,ArkaniHamed:1998nu,Dudas:2007nz,Gallego:2008sv,Dudas:2008qf}.
A brief review of models with these characteristics is done in
appendix \ref{appLVSwithMatter}.

\subsection*{$N_F<N_C$ case}

In the window $N_F<N_C$, with $N_f$ the number of families and
$N_C$ the rank of the gauge group, that we suppose to be an
$SU(N_C)$, the system develops an ADS superpotential
\cite{Affleck:1983mk} which in the particular case $N_F=1$ and
$N_C=2$ takes the following form
\begin{equation}
W=W_{cs}+A\frac{e^{-a t}}{\phi^2}+\frac12m \rho \phi^2\,,
\end{equation}
where the chiral matter fields charged under the non-Abelian
sector are described by the canonical normalized meson
$\phi=\sqrt{2 Q\tilde Q}$. As before $W_{cs}$ and $A$ are ${\cal
O}(1)$ functions of $H$ fields. We have also added a mass term, in
general $H$-dependent, which from gauge invariance under the
Abelian sector depends on a second matter like field $\rho$. We
normalize the charges such that $\delta^t=2/a$, thus $q_\phi=-1/2$
and $q_\rho=1$. This model although proposed in
\cite{Cremades:2007ig} was not studied there, so we take the
opportunity to carefully check if indeed leads to a LVS and
characterize its vacua (see appendix \ref{appLVSwithMatter}).
Despite the fact it turns out that as it stands it cannot lead to
realistic solutions with vanishing cosmological constant we do the
analysis of the decoupling having the model interesting
characteristics like the mass term in the superpotential. The
modular weights for both matter fields are taken equal and the
kinetic gauge function is taken to be $f_X\sim t$. The VEV of
$\phi$ is ruled by the cancellation of the $D$-term dynamics,
meanwhile for $\rho$ a non-vanishing VEV is only possible for a
non-zero mass parameter. The moduli are ruled by a potential whose
structure is similar to the standard LVS but with different powers
of the volume. Nicely enough a fully analytical solution is found
in the large volume expansion, with out-shot,
\begin{equation}
e^{-a\,t}\sim \epsilon^{2-n}\,,~~t\sim 1\,,~~\phi\sim
\epsilon^{\frac{1-n}{2}}\,,~~\rho\sim\epsilon^{\frac{5-3n}{6}}\,.
\end{equation}
Contrary to the first two cases, here it is not necessary to argue
for a tuning in the GS coefficient, however, the mass term may
lead the dynamics and sweep the solution, so should be tuned by
$m\sim \epsilon^{\frac{3n-1}{2}}$. With this consideration the
following scalings are found,
\begin{align}
G_\alpha\sim
\Big(\epsilon^{\frac23},\epsilon,\epsilon^{\frac{1+n}{2}},\epsilon^{\frac{1+n}{2}}\Big)\,,&~~
G_{\alpha \bar\beta}\sim \begin{pmatrix}
\epsilon^{4/3} & \epsilon^{5/3} &\epsilon^{\frac{7+3n}{6}}&\epsilon^{\frac{3+n}{2}}  \\
\epsilon^{5/3}&  \epsilon & \epsilon^
{\frac{1+n}{2}} & \epsilon^{\frac{5+3n}{6}} \\
\epsilon^{\frac{7+3n}{6}}  & \epsilon^
{\frac{1+n}{2}} & \epsilon^{n} & 0\\
\epsilon^{\frac{3+n}{2}} & \epsilon^{\frac{5+3n}{6}}  & 0 &
\epsilon^{n}
\end{pmatrix}\,,
\end{align}
with ordering $\{T,t,\phi,\rho\}$, and where for simplicity in the
analysis, contrary to the previous examples, we have replaced the
matter fields by its VEV to extract the scalings.  The mixed
derivatives scale as
\begin{align}
G_{i\alpha}\sim
\Big(\epsilon^{5/3},\epsilon\,,\epsilon^{\frac{1+n}{2}}\,,\epsilon^{\frac{1+n}{2}}\Big)\,,&~~
G_{i\alpha\bar \beta}\sim
\begin{pmatrix}
\epsilon^{7/3}               & \epsilon^{8/3} & \epsilon^{\frac{7+3n}{6}} &\epsilon^{\frac{3+n}{2}}  \\
\epsilon^{8/3}               & \epsilon^2     & \epsilon^
{\frac{1+n}{2}} & \epsilon^{\frac{5+3n}{6}} \\
\epsilon^{\frac{7+3n}{6}}    & \epsilon^
{\frac{1+n}{2}} &\epsilon^{n}& 0\\
\epsilon^{\frac{3+n}{2}}  &\epsilon^{\frac{5+3n}{6}}  & 0 &
\epsilon^{n}
\end{pmatrix}\,,
\end{align}
from which we estimate the VEV of the auxiliary fields,
\begin{equation}
F^T\sim \epsilon^{1/3}\,,~~F^t\sim\epsilon\,,~~F^\phi\sim
\epsilon^{\frac{3-n}{2}}\,,~~F^\rho\sim
\epsilon^{\frac{3-n}{2}}\,~~ \mbox{and}~~F^i\sim\epsilon^2.
\end{equation}
We have also $G_X\sim D^X\sim \epsilon^2$. The canonical
normalized auxiliary fields VEV's, encoding the SUSY breaking
contribution from each sector, scale as $F^T_c\sim \epsilon$,
$F^t_c\sim \epsilon^{3/2}$, $F^\phi_c\sim \epsilon^{3/2}$,
$F^\rho_c\sim \epsilon^{3/2}$ and $F^i_c\sim \epsilon^2$. The same
analysis followed before shows corrections to the simplified
version suppressed by ${\cal O}(\epsilon^{2/3})$. Nicely enough,
even though we are now dealing with a fourth suppression factor,
namely the mass parameter, the leading order for the corrections
to the simplified version are again independent of the modular
weight. Like in the other cases, this is due to the tuning on the
parameters from the requirement of a vanishing cosmological
constant, or more weakly, the requirement that the stabilization
of the moduli is not spoiled. A related result is the independence
on the modular weight of the VEV of non-pertubative
superpotential, being in all cases ${\cal O}(\epsilon)$, and even
the mass term in this last example goes like $m \rho \phi^2\sim
\epsilon^{4/3}$.
\\
The largest corrections we are spotting here come from
$G_{i\rho}\sim \phi^2
\partial_i m\sim \epsilon^{\frac{1+n}{2}}$ together with $G_{iT}$ and
$G_{it}$ affecting the $G_{T\bar\rho}F^TF^{\bar
\rho}+G_{t\bar\rho}F^tF^{\bar \rho}+h.c.$ terms in the simplified
version. Therefore, in case the mass parameter is absent, or
independent of the $H$ fields, the corrections are smaller and
turn out to be like in the previous cases ${\cal O}(\epsilon)$. In
the numerical check done in appendix \ref{NumTest}, for example,
it turns out that the ${\cal O}(\epsilon)$ corrections are as
important as the ones coming from the mass parameter due to
enhancement factors like the ones found before.

\subsection*{$N_C< N_F<3/2 N_F$ case}

A realistic model already present in the literature is the
Krippendorf-Quevedo model developed in \cite{Krippendorf:2009zza},
regarding the Seiberg dual description of the theory in the window
$N_C< N_F<3/2 N_F$. The model can be minimally described by two
matter fields with the following superpotential,
\begin{equation}
W=W_{cs}+ A e^{-a\, t}\rho \phi\,,
\end{equation}
and as before $W_{cs}$ and $A$ can depend on the $H$ fields and
are regarded to be ${\cal O}(1)$.\footnote{See appendix
\ref{KQModel} where is argued that such assumption breaks down
whenever the volume turns out to be huge. Here we simply disregard
such subtlety which makes obscure the analysis.} The charges can
be normalized fixing the GS coefficient to be $\delta^t=2/a$ so
that, $q_\phi=-1$ and $q_\rho=2$. The same modular modular weight
is taken for both matter fields and the kinetic gauge function is
taken to be $f_X\sim t$. Like in the previous case the VEV of
$\phi$ is ruled by the $D$-term minimization. A rough analytical
approach shows the scalings to be
\begin{equation}\label{VEVKPmodel}
e^{-a\,t}\sim \epsilon^{n+1/2}\,,~~t\sim 1\,,~~\phi\sim
\epsilon^{(1-n)/2}\,,~~\rho\sim \epsilon^{(2-n)/2}\,.
\end{equation}
Then for the analysis we have the scalings
\begin{align}
G_{\alpha}\sim
&\Big(\epsilon^{2/3},\epsilon,\epsilon^{(1+n)/2},\epsilon^{(2+n)/2}\Big)\,,&~~
G_{\alpha\bar\beta}\sim
\begin{pmatrix}
\epsilon^{\frac{4}{3}}     & \epsilon^{\frac{5}{3}}    &\epsilon^{\frac{7+3n}{6}} & \epsilon^{\frac{10+3n}{6}} \\
\epsilon^{\frac{5}{3}}     &  \epsilon                 & \epsilon^{\frac{1+n}{2}} & \epsilon^{\frac{2+n}{2}} \\
\epsilon^{\frac{7+3n}{6}}  & \epsilon^{\frac{1+n}{2}}  & \epsilon^{n}             &0\\
\epsilon^{\frac{10+3n}{6}} & \epsilon^{\frac{2+n}{2}}  & 0 &
\epsilon^{n}
\end{pmatrix}\,,
\end{align}
with ordering $\{T,t,\phi,\rho\}$ and where like in the previous
case we replaced explicitly the VEV of the matter fields so to
make simpler the analysis. In the vacuum also $D^X\sim G_X\sim
\epsilon^2$, which together with the mixed derivatives with the
$H$ sector
\begin{align}
G_{i\alpha}\sim
&\Big(\epsilon^{\frac53},\epsilon,\epsilon^{\frac{1+n}{2}},\epsilon^{\frac{2+n}{2}}\Big)\,,&~~
G_{i\alpha\bar\beta}\sim
\begin{pmatrix}
\epsilon^{\frac{4}{3}}     & \epsilon^{\frac{5}{3}}    &\epsilon^{\frac{7+3n}{6}} & \epsilon^{\frac{10+3n}{6}} \\
\epsilon^{\frac{5}{3}}     &  \epsilon                 & \epsilon^{\frac{1+n}{2}} & \epsilon^{\frac{2+n}{2}} \\
\epsilon^{\frac{7+3n}{6}}  & \epsilon^{\frac{1+n}{2}}  & \epsilon^{n}             & 0\\
\epsilon^{\frac{10+3n}{6}} & \epsilon^{\frac{2+n}{2}}  & 0 &
\epsilon^{n}
\end{pmatrix}\,,
\end{align}
imply
\begin{equation}
F^T\sim \epsilon^{1/3}\,,~~F^t\sim \epsilon\,,~~F^\phi\sim
\epsilon^{\frac{3-n}{2}}\,,~~F^\rho\sim
\epsilon^{\frac{4-n}{2}}\,,~~F^i\sim \epsilon^2\,.
\end{equation}
The analysis follows like in the previous cases finding
corrections suppressed at least by ${\cal O}(\epsilon)$, lead by
the equivalent ones found in the pure moduli case; indeed, the VEV
for the non-perturbative part of the superpotential again scales
like ${\cal O}(\epsilon)$ and the corrections turn out to be
independent of the modular weight. Again the canonical normalized
$F$-terms: $F^T_c\sim \epsilon$, $F^t_c\sim\epsilon^{3/2}$,
$F^\phi_c\sim \epsilon^{3/2}$, $F^\rho_c\sim \epsilon^2$ and
$F^i_c\sim\epsilon^2$, show the suppressed contribution from the
$H$ sector to the breaking of SUSY. Unfortunately the rough
analytical approach we followed in finding the solutions is not
enough for our computer skills so to allow us to find numerically
this kind of vacua and check the conclusions just raised.

\subsection*{Higher order operators and universality in the corrections}

So far we have regarded only the lowest order operators in the
K\"ahler potential neglecting contributions of the form $K\supset
Q^p/{\cal V}^m$ with $p>2$. Although the introduction of such
terms implies to deal with extra suppression factors, as the power
$m$ is in principle unrelated to the modular weight $n$, we can
from our results easily spot the effects of integrating out the
$H$ fields in such a case. In section \ref{ONLYCHIRALVSsect} we
found that the simplified version misses terms that are suppressed
by ${\cal O}(\epsilon ^n Q^2)$, where the $Q^2$ can be seen as
fluctuations of the matter fields. This means that the full
effective theory realizes terms like ${\cal L}_{eff}\supset
\epsilon^{2+2n}Q^4$, which affect terms coming from the higher
coupling in the K\"ahler potential with $p=4$, namely ${\cal
L}_{simp}\supset \epsilon^{2+m} |Q|^4$, e.g., from $G_{Q\bar
Q}|F^Q|^2$, telling us that such a term in the K\"ahler potential
is not reliable unless $m< 2n$. Similar comments follow for any
other higher couplings. These higher order effective terms can be
easily understood by collapsing Feynman diagrams with $H$ fields
in the internal lines, finding also the same kind of corrections
once higher order operators are regarded in the superpotential;
then, a ${\cal O}(1)$ Yukawa coupling can induce operators
correcting the $Q^6$ operator in the superpotential and the $Q^5$
one in the K\"ahler potential, exactly as is found even in case a
mass hierarchy is realized \cite{Gallego:2009px}.
\\
The above discussion is somehow related to the fact that contrary
to the moduli, the derivatives with respect the matter fields
always carry a decreasing on the suppression factor controlling
the factorization of the theory, since these fields scale with
inverse powers of the volume. Then, at some order in the matter
fields fluctuations the factorizability is expected to breakdown
and the corrections start to be non-suppressed.
\\
Notice that this analysis points out also, and contrary to claim
done above, for the possibility the corrections to the lower order
operators to be dependent on the modular weights. Indeed, the
corrections of ${\cal O}(\epsilon^n Q^2)$ can be seen also as
${\cal O}(\epsilon^n Q)\delta Q$ corrections with $\delta Q$ the
fluctuations. Let us see more precisely that in fact up to the
mass level, at least for the models studied, the corrections are
universal.
\\
For this we write down the scalar potential expanding in
fluctuations $\delta Q$ and $\delta H$ around the $H_o$ solution,
\begin{align}
V&=V_o+V_{i}\delta H+V_Q\delta Q+V_{Qi}\delta Q\delta H+M_H^2
\delta H \delta H+V_{QQ}\delta Q\delta Q\cr &\sim
\epsilon^3+\epsilon^3\delta H+\epsilon^{\frac{5+n}{2}}\delta Q
+\epsilon^{\frac{5+n}{2}}\delta Q\delta H+\epsilon^2 \delta
H\delta H+\epsilon^{2+n}\delta Q\delta Q\,,
\end{align}
where we have used the scalings at the vacuum, and the fact that
$V_{Qi}\sim V_Q\sim QV_{QQ}$, valid for the fields not stabilized
by the $D$-term cancellation in which case a larger suppression
appear in the $V_Q$ and therefore things are even safer. We also
used that the non-canonical mass squared for these matter fields
turn out to be always of ${\cal O}(\epsilon^{2+n})$ and the
largest VEV for the matter is ${\cal
O}(\epsilon^{\frac{1-n}{2}})$. An expansion in the moduli
fluctuation is also implicit in the coefficients but for our
present purposes are not relevant. Then an integration at the
gaussian level leads to $\delta H\sim
\epsilon+\epsilon^{\frac{1+n}{2}}\delta Q$, and plugging these
back in the potential we have corrections of the form
\begin{equation}
\delta V=\epsilon^4+\epsilon^{\frac{7+n}{2}}\delta
Q+\epsilon^{3+n}\delta Q\delta Q\,,
\end{equation}
which are suppressed compared to the original operators by ${\cal
O}( \epsilon)$, independent of $n$. These as far the largest
corrections are concerned, since we took the largest VEV for the
$Q$'s. This analysis, in fact, is quite general as uses the
largest possible VEV for a generic $Q$ field, in an arbitrary LVS
model, therefore the conclusion is expected to hold in a more
general context.

\section{Conclusions}\label{ConcluSect}

Our results show how the factorizable models present a framework
where freezing of a set of light fields is a reliable procedure.
Although several papers have worked on these models
\cite{Achucarro:2007qa,Achucarro:2008sy,Achucarro:2008fk,Achucarro:2010da},
we present a generalization which describes the introduction of
matter like fields, whose wave function depends in general on
fields from both sectors. Then, once these fields enter only in
the suppressed part of the generalized K\"ahler function,
eq.\eqref{GENFACT}, decoupling between the two sectors is easily
understood through the same analysis done in the pure moduli case,
with the same conclusion: freezing of the $H$ fields is reliable
as far as the $H$ lowest components stand at their leading
$F$-flatness solutions. The corrections, however, contrary to the
pure moduli case, where start at the next-to-next-to-leading order
in the suppression parameter $\epsilon$, start now at the
next-to-leading order. With gauge interactions turned on this
conclusion does not change but one has to require the $H$ fields
to be neutral and, in case the $D$-term SUSY breaking be non
suppressed, that the gauge kinetic function dependency on these
fields be suppressed. These models although not realistic, with a
SUSY breaking scale or order the Planck mass, present all the
important features needed to understand how the freezing of the
Dilaton and Complex structure moduli proceeds in the LVS of
type-IIB string compactifications. Nicely enough, in these
scenarios the $D$-term SUSY breaking is naturally suppressed by
some powers of the volume and the constrain found for the gauge
kinetic function is avoided.
\\
A somehow unexpected result is that, in all models explicitly
studied, the largest corrections to the simplified version, up to
the mass level in the matter fluctuations, are independent of the
modular weight for the matter fields. This due to the tuning on
the parameters and a constrain on the largest possible VEV a $Q$
field can get, both related to the requirement of a vanishing
cosmological constant, or the more relaxed and general one of
avoiding the destabilization of the vacuum. On the other hand,
higher order corrections may depend on the modular weights. These,
however, are expected to be irrelevant for moduli stabilization
issues and therefore harmless for the reliability of the Two-Step
procedure.
\\
Our study for the LVS, although far from being fully generic,
shows how the Two-Step stabilization procedure is expected to be
safe in realistic and complicated scenarios, and how the
corrections are universally independent of the modular weights. It
also introduces a simple way to estimate the mistake done by
performing quantitative analysis from the simplified version.
Still, a more conservative approach can use the fact that in
moduli stabilization studies the harder part for the quantitative
information is finding the vacua, say numerically. Then, once the
vacuum is found through the simplified version one can implement
the full model knowing that the solutions are just slightly
perturbed. This, in fact, is the procedure followed in the
analytical and numerical examples shown in the appendices. With
the full information at hand one is safe of missing fine
information probably needed for precision tests, for example soft
terms ruled by the $H$ fields. Off course, explicit string
implementations 
most likely would need a sort of
Two-Step procedure, even if less drastic, due to the huge number
of fields. Nevertheless, our study sets down a consistency check
for the LVS as a systematic procedure for moduli stabilization.
\\
An interesting result our analysis shows, is the SUSY nature of
the effective theory despite the lack of a hierarchy between the
energy scales in both sectors, in particular with the SUSY
breaking scale. One can understand this from the fact that the
main requirement to be satisfied in order the effective theory be
approximately SUSY, is that the solution for the auxiliary fields
of the multiplets being integrated out be always suppressed
compared with the auxiliary fields of the remaining fields,
independently of the dynamics of the effective theory. One way of
realizing such situation, as found by Brizi et al.
\cite{Brizi:2009nn}, is by integrating out fields with large SUSY
mass compared to the SUSY breaking scale, driven this last one by
the remaining dynamics. In this case is the large inertia from the
SUSY preserving sector what shields it from the SUSY breaking
effects in the second sector, and the auxiliary field VEV's of
these multiplets are suppressed by the masses. In our case, is the
weak coupling among the different sectors what ensures that the
solution in one side be independent of the other one. Then any
back reaction from the SUSY breaking sector is negligible in the
SUSY preserving sector. As a matter of fact, in case the mixing
parameter $\epsilon$ vanishes the solution for the $F^i$ auxiliary
fields exactly vanishes and one can even speak about an exact SUSY
description of the effective theory, contrary to the case with the
hierarchy where there are always corrections to the two derivative
SUSY description of the effective theory \cite{Brizi:2009nn}. In
order to put these arguments in a complete SUSY framework, one
should understand them from a fully superfield approach, a study
left for a forthcoming paper \cite{Inprep}.

\acknowledgments{I would like to thank Marco Serone for important
discussions in the beginning of the work, for a careful reading of
a preliminary version of the manuscript and his relevant comments
on it. I thank also Fernando Quevedo for encouraging comments
about the work.}

\appendix

\section{LVS with matter}\label{appLVSwithMatter}

This appendix is devoted to resume the principal features of the
LVS models with matter and gauge interactions, for details please
refer to the original works
\cite{Cremades:2007ig,Krippendorf:2009zza}.\footnote{There is also
an interesting realization of LVS in Heterotic compactifications
recently developed by L. Anguelova and C. Quigley in
\cite{Anguelova:2010qd}. However, the phenomenological constrain
on the Dilaton VEV, $\langle S\rangle\sim 2$ coming from an good
approximate value of the gauge couplings, together with the
expression for the string coupling $g_s\sim \sqrt{T^3/S}$
\cite{Witten:1985xb}, translates to a tough constrain on the VEV
of the volume as the theory starts now to leave the perturbative
regime.} In the following we use the more orthodox form for the
scalar potential integrating out the full set of auxiliary fields,
\begin{equation}
V=e^G\big( G_{\bar\alpha}G^{\bar\alpha
\alpha}G_{\alpha}-3\big)+\frac{1}{2h_X}(G_X)^2\,,
\end{equation}
with the matrix $G^{\bar\alpha\alpha}\equiv G_{\alpha\bar
\alpha}^{-1}$, and the rest of the functions as defined in the
main text. We restrict the study to the simplified version of the
models, where the Dilaton and Complex structure moduli are
regarded as frozen, since our conclusions show that all features
we need are encoded here. Moreover we neglect the now constant
part ${\cal K}_{cs}$ in the K\"ahler potential,
eq.\eqref{ModKPot}, changing only by a multiplicative factor the
whole $F$-term scalar potential, and as before we neglect the
non-Abelian parts in the $D$-term dynamics.
\\
The K\"ahler moduli enter in the superpotential only though
non-perturbative effects, exponentially suppressed by the moduli,
then any dependency on the large modulus can be neglected. In
absence of matter fields the superpotential then looks like
\begin{equation}\label{trivialW}
W=W_o+Ae^{-a\,t}\,,
\end{equation}
with $W_o$ the ${\cal O}(1)$ remnants of a flux induced
superpotential once the Dilaton and Complex structure are frozen,
i.e., $W_o\equiv W_{SC}(U^i_o,S_o)$, and the amplitude $A$ might
also depend on these moduli. The standard scenario works without
gauge interaction nor matter like fields, then the scalar
potential, in a large volume expansion and regarding $t$ and all
the parameters as real, reads \cite{Balasubramanian:2005zx},
\begin{equation}\label{standarLVSpot}
V_F\approx \frac{8}{3\lambda}\frac{a^2\sqrt{t}|A|^2}{{\cal
V}}e^{-2a\,t}-4a\,t\frac{|A\,W_o|}{{\cal
V}^2}e^{-a\,t}+\frac{3}{2}\frac{\hat\xi |W_o|^2}{{\cal V}^3}\,,
\end{equation}
where $\hat \xi=\xi (2 S_{r})^{3/2}$. The competing effects of
these three terms lead to a stable minimum, being crucial that the
last term be positive definite, namely $\hat\xi>0$, which
translates to a condition on the kind of CY, in a generic setup,
where such minima can exist, i.e., as $\xi\sim h^{2,1}-h^{1,1}$
this implies more complex structures than K\"ahler moduli
\cite{Cicoli:2008va}. The minimization of this potential leads to
\cite{Balasubramanian:2005zx}
\begin{equation}
t\approx \left(\frac{\hat \xi}{\lambda}\right)^{2/3}\,,~~{\cal
V}\approx \frac{9\lambda|W_o|}{4a|A|}\sqrt{t}e^{at}\,,
\end{equation}
with corrections of order $1/at\ll 1$. The canonical normalized
masses $m_T\sim \epsilon^{3/2}$ and $m_t\sim \epsilon$ can be
estimated by disregarding the K\"ahler mixing, i.e., $m_i^2\sim
(K_{ii})^{-1}\partial_i\partial_i V$, using the notation ${\cal
V}^{-1}\sim \epsilon$ as in the main text. Notice the possibility
of realizing low energy SUSY breaking, characterized by the
gravitino mass $m_{3/2}=e^{G/2}\sim \epsilon$, with a natural VEV
for $W$ driven by the exponentially sized volumen. The problem
with this setup is the fact that the vacuum turns out to be a deep
$AdS$, of ${\cal O}(\epsilon^3)$, as can be seen from the fact
that the second term in \eqref{standarLVSpot} has in the amplitude
a factor $at\gg 1$ which enhances it compared to the other two.

\subsection{Charged large modulus}\label{appNoQinW}

Once matter fields are considered, with a K\"ahler potential like
the one in eq.\eqref{KahlerQ}, even if these turn out not to
appear in the superpotential the $F$-term scalar potential is
affected with new terms lead, in a large volume expansion, by
\begin{equation}\label{uplift1}
V_F\supset e^G G^{Q\bar Q}|G_Q|^2\sim e^{G}G_{Q\bar
Q}^{-1}|G_Q|^2\sim \epsilon^{2+n}|Q|^2\,.
\end{equation}
If such terms are not to sweep off the minimum of the moduli,
expected to proceed like in the standard scenario, and potentially
realize a zero tuned cosmological constant, these should scale at
most as $\epsilon^3$, so we get a constrain on the VEV for the
$Q$'s,
\begin{equation}
\langle|Q|^2\rangle\lesssim \epsilon^{1-n}\,.
\end{equation}
A superpotential like eq.\eqref{trivialW} still is possible with a
neutral small modulus, and the appearance of the $Q$ fields might
be forbidden by gauge invariance, then the generic form for the
$D$-term part of the scalar potential is the one give in
eq.\eqref{VDpot}.

\subsubsection*{Vanishing $Q$ VEV}\label{VanishingQVEVmodel}
The simplest and nicely enough also a realistic scenario
developing vanishing cosmological constant, is the one where the
$Q$ charge has opposite sign of the GS coefficient, so that all
terms in the $D$-term are positive definite, as well the mass term
in the $F$-term potential. Then a vanishing VEV solution for $Q$
is possible.
\\
The uplift, thus, is not possible through \eqref{uplift1} but from
the $D$-term potential scaling as
\begin{equation}
V_D\sim\frac{1}{h_X}\left(\frac{\delta^T}{T+\bar
T}\right)^2\sim\epsilon^{4/3}\frac{\big(\delta^T\big)^2}{h_X}\,,
\end{equation}
so that $G_X\sim \delta^T\epsilon^{2/3}$ and $D^X\sim
\delta^T\epsilon^{2/3}/h_X$. Then in order this to work as an
uplift for the $AdS$ solution of $V_F=-{\cal O}(\epsilon^3)$, and
no to spoil the moduli stabilization, we find that the $GS$
coefficient should be tuned as $\delta^T\lesssim \epsilon^{5/6}
\sqrt{h_X}$, and even in the best case when $h_X\sim
T\sim\epsilon^{-2/3}$ we have a strong constrain $\delta^T\lesssim
\epsilon^{1/2}$.
\\
Contrary to the standard scenario no massless modes are realized,
as the imaginary part of $T$ is the would-be Goldstone boson eaten
up by the now massive vector field of the broken $U(1)_X$. For the
$Q$ field the mass is ruled by the term \eqref{uplift1}, so that
the canonical mass goes like $m_Q\sim \epsilon$. The gauge
breaking scale, encoded in the canonical mass for the vector
bosons, is given by $m_X=\sqrt{K_{T\bar T}X^T\bar X^{\bar
T}/h_X}\sim \epsilon^{2/3}\delta^T/\sqrt{h_X}\sim \epsilon^{3/2}$.

\subsubsection*{Non-vanishing $Q$ VEV}\label{nonzeroQVEVmodel}

When the sign of $q_Q$ is opposite to the one of the GS
coefficient the minimization of the potential leads in a good
approximation to a cancellation of the $D$-term potential, driven
by a non-vanishing $Q$ VEV, so
\begin{equation}
\langle |Q|^2\rangle\approx \frac{\delta^T}{q_Q}\frac{{\cal
V}^n}{Z(t_r) T_r}\sim \epsilon^{2/3-n}\delta^T\,.
\end{equation}
Let us see the structure for the full e.o.m. for $Q$, so to have a
more precise estimation for the VEV of the $D$-term part of the
potential
\begin{align}\label{FULLEOMFORQ}
\partial_QV\approx \partial_Q \Big(e^{G}G_{Q\bar Q}^{-1}|G_Q|^2\Big)+\frac{1}{2h_X}G_XG_{QX}\sim\epsilon^{2+n}Q+\frac{1}{h_X}\epsilon^n Q
G_X=0\,,
\end{align}
where without loss of generality we regard $h_X$ independent of
$Q$, the discarded term being of order $(G_X)^2$ is any way
irrelevant. Then, we find that
\begin{equation}
G_X\sim h_X\epsilon^2\,,~~D^X\sim G_X/h_X\sim \epsilon^2\,.
\end{equation}
We have, then, that the $D$-term part of the potential scales as
$V_D\sim h_X \epsilon^4$ and even in the best case
$h_X\sim\epsilon^{-2/3}$ it can not work as uplifting for the
moduli vacuum which follow again like in the pure moduli case.
This role is now played by the contribution \eqref{uplift1},
scaling as
\begin{equation}
V_{uplift}\sim\epsilon^{2+n}|Q|^2\sim\epsilon^{8/3}\delta^T\,.
\end{equation}
Again a tuning on the GS coefficient, $\delta^T\lesssim
\epsilon^{1/3}$, is required though in this case is milder and
furthermore independent of the scaling of $h_X$.
\\
The physical mass scalings follow exactly like in the previous
case, with an important contribution to the mass of $Q$ from the
$D$-term dynamics though, and the canonical normalized mass for
the vector boson $m_X\sim\epsilon^{1/2}/\sqrt{h_X}$.

\subsection{Charged small modulus}

Things are more reach when the small modulus is charged as the the
superpotential starts necessarily to depend on the $Q$ fields. As
is well known depending on the number of families and the rank of
the gauge group the description of the system changes. 

\subsubsection*{$N_f<N_c$ case}\label{appNewModel}

The following model was already proposed by Cremades  et al. in
\cite{Cremades:2007ig}, however, it was not discussed there. Here
we discuss the stabilization of the fields in detail and try to
argue why is not possible to realize Minkowski/dS vacua with this
setup.
\\
Working in the window $N_f<N_c$ for a non-Abelian $SU(N_c)$ sector
an $ADS$ nonperturbative superpotential \cite{Affleck:1983mk} is
generated, depending on the non-perturbative scale $\Lambda$ and
the chiral fields, $Q$ ($\tilde Q$), transforming in the
fundamental (antifundamental). It is convenient to work in
$D$-flat directions characterized by the mesonic fields, then
choosing the canonical normalized field $\phi=\sqrt{2 Q\bar Q}$
\cite{ArkaniHamed:1998nu}, and regarding each family as the copy
of the others the non-perturbative superpotential takes the form
\begin{equation}
W_{np}=(N_c-N_f)\left(\frac{2\Lambda^{3N_c-N_f}}{\phi^{2N_f}}\right)^{\frac{1}{N_c-N_f}}\,.
\end{equation}
The non-perturbative scale is a function of the moduli which as
before we take to be the small modulus. To make easier the
following analysis we take the simplest case, namely $N_c=2$ and
$N_f=1$, then the superpotential takes the general form,
\begin{equation}
W=W_o-A \frac{e^{-a\, t}}{\phi^2}-\frac12  m \rho \phi^2\,,
\end{equation}
where we have included the flux induced superpotential that
stabilizes the Dilaton and Complex structure moduli, and encoded
some $N_c$ and $N_f$ dependencies in the parameters $A$ and $a$
that are naturally of order one. A possible mass term is also
included, which requires a further charged field, $\rho$, singlet
under the $SU(2)$.\footnote{The phase between the terms is
irrelevant as can be modified by the value of the coefficient and
have been fixed only in order to clear up the following formulae
meanwhile the coefficients are regarded as positive.}
\\
Without loss of generality we fix the charge for the modulus by
$\delta^t=2/a$, so the holomorphic Killing vectors are
$X_X=i(0,1/a,-\phi/2,\rho)$. The K\"ahler potential for the matter
fields is taken  to be $K\supset \frac{Z(t_r)}{T_r^n}|Q|^2$, so
the leading expression for the $D$-term
\begin{equation}
G_X=-iX^iG_i\approx\frac12\left(\frac{1}{a}\frac{3\sqrt{t_r}}{T_r^{3/2}}+\frac{Z(t_r)}{T_r^n}(2\rho\bar\rho-\phi
\bar \phi)\right)\,,
\end{equation}
where we have also discarded subleading terms in $1/at$.
\\
Despite the mass term for the mesons these cannot be integrated
out since the tachionic mass from the $D$-term dynamics likely
dominates. In fact, the minimization of the potential leads to a
cancellation of the $D$-term potential via a non-vanishing VEV for
$\phi$, analogous to the previous case, but now this happens for
the opposite sing of the charge as the sign of the field dependent
FI term changes. Then $\langle \phi^2\rangle\approx
\frac{3\sqrt{t_r}}{aZ(t_r)T_r^{3/2-n}}\sim\epsilon^{1-2n/3}\phi_o(t_r)^2$,
taking {\it a priori} $\langle \phi \rangle\gg \langle
\rho\rangle$ and $\phi_o(t_r)$ encoding factors scaling like
${\cal O}(1)$. In order to stress the fact that the dynamics from
the mass term are weaker let us discuss first the case $m=0$ and
then see the consequences of turning this term on. Thus, taking
the gauge kinetic function $f_X=t$ the leading $D$-term potential
for $\rho$ once $\phi$ is fixed is
\begin{equation}\label{rhoDterm}
V_D\approx\frac{Z(t_r)^2}{t_r T_r^{2n}}|\rho|^4\,.
\end{equation}
From here on we consider real solutions for the e.o.m. and forget
the subscript $r$, this is later checked numerically for real
parameters. Linear terms in $\rho$ would appear only as factors of
the mass term with leading expression, coming from the
$G^{\phi\bar\phi}|G_\phi|^2$ and $G^{t\bar \phi}G_tG_{\bar
\phi}+h.c.$ terms, which once $\phi $ is fixed reads at leading
order
\begin{equation}
-\frac{8}{\lambda^2
Z(t)T^{3/2}}m\,\rho\left(\frac{t\,W_oZ^\prime(t)\phi_o(t)^2}{4T^{3-n}}+4a\,A
e^{-a\,t}\big(\phi_o(t)^{-2}-a\,t Z^\prime(t)\big)\right)\,.
\end{equation}
Therefore, for the case we are currently studying a solution
$\rho=0$ is possible. The mass terms for $\rho$ are, at leading
order, given by
\begin{equation}\label{rhocua}
\frac{4m^2\phi_o(t)^2}{\lambda^2 Z(t)T^{9/2-2n}}-\frac{8a t A
e^{-at}}{9\lambda^2Z(t)\phi_o(t)^4}\left(\frac{3\sqrt{t}\,W_o
\phi_o(t)^2 }{Z(t)^2T^{2n}}-2 a A e^{-a\,t}
\right)(Z^\prime(t)^2-Z(t)Z^{\prime\prime}(t))\,,
\end{equation}
again lead by the $G_t$ and $G_\phi$ terms. With this fixed
solutions for $\rho$ and $\phi$ we get an effective potential for
the moduli, which at leading order in the volume and $1/at\ll 1$
reads,
\begin{align}\label{ModPot}
V_{mod}=&\frac{8 Z(t)^2 a^4 A^2 e^{-2a\,t}}{27\lambda^2 \sqrt{t}
T^{2n-3/2}}-\frac{2\sqrt{t}Z(t)a^2 W_o A e^{-a\,t}}{\lambda^2
T^{3/2+n}}+\frac{3 W_o^2\hat \xi}{2\lambda^3 T^{9/2}}\,,
\end{align}
where the first two terms come from the $G^{t\bar t}|G_t|^2$
contribution and the last one is the usual $\alpha^\prime$ term
from $G^{T\bar T}|G_T|^2$. Interestingly enough the structure of
this potential is exactly like the standard LVS,
eq.\eqref{standarLVSpot}, with a change in the powers for the
large modulus. This change in the powers induces a different
scaling for the exponential since the minimization leads to the
same scaling of all three terms at the minimum. Indeed we see that
at the minimum the exponential will in this case scale like
$e^{-a\,t}\sim T^{n-3}\sim \epsilon^{2-2n/3}$, while in the
standard scenario $e^{-a\,t}\sim T^{-3/2}\sim\epsilon$. In order
to have cleaner formulae in the following we fix $Z(t)=t$, then
the minimization in $T$ leads to a solution of the form, using
$a\,t\gg \lambda$,
\begin{equation}
\langle T\rangle \approx \left\{\frac{27 (3+2n)W_o
e^{a\,t}}{8(4n-3)a^2 A}\left(1-\sqrt{1-\frac{4(4n-3)\hat \xi
}{(3+2n)^2t^{3/2}\lambda}}\right)\right\}^{\frac{1}{3-n}}\approx
\left(\frac{27 W_o e^{a\,t}}{4(3+2n)\lambda a^2A t^{3/2}}
\right)^{\frac{1}{3-n}}\,.
\end{equation}
Plugging this solution into the e.o.m. for the small modulus lead
by the derivatives of the exponential factors one gets the
following real solution
\begin{equation}
\langle t\rangle\approx\left(\frac{a\hat
\xi}{(3+2n)\lambda}\right)^{3/2}\,,
\end{equation}
a very similar result to the one of the standard LVS.
\\
At the true minimun the $D$-term actually is not zero as the
$F$-term part also contributes to the e.o.m for $\phi$. The
estimation for its correct VEV is done by refining the e.o.m. for
$\phi$ including the leading contribution from the $F$-term
potential, analogous to eq.\eqref{FULLEOMFORQ}. This shows that at
the vacuum $G_X\sim \epsilon^2$ and so it is $D^X$.
\\
Like in the standard scenario the vacuum, so far, turns out to be
an $AdS$ one, again of ${\cal O}(\epsilon^3)$, since the negative
term in \eqref{ModPot} leads in a $at\gg 1$ expansion, and the
$D$-term dynamics, although positive definite cannot do the job
since scales like ${\cal O}(\epsilon^4)$.
\\
When a non-vanishing value for the mass parameter is turned on not
only a non-vanishing VEV for $\rho$ is generated but also a term
of the form,
\begin{equation}
V_F\supset \frac{1}{4Z(t)T^{3-n}}m^2 |\phi|^4\,,
\end{equation}
coming from $G^{\rho\bar\rho}|G_\rho|^2$, which being positive
definite can potentially uplift the vacuum this as far there is
tuning in the mass parameter of order $m\sim \epsilon^{n-1/2}$.
Notice that the term \eqref{rhoDterm} might work also since now
the VEV of $\rho$ is non zero. However, the minimization of
$\rho$, driven by \eqref{rhoDterm} and the linear term leads to
$\rho\sim \epsilon^{5/6-n/3}$, so that \eqref{rhoDterm}, and the
linear term also, scale like $\epsilon^{10/3}$ not enough for the
uplift.
\\
Unfortunately by numerical proof we found that the perturbation on
the vacuum once the mass term is turned on does not allow to
uplift and the minimum continue to be an AdS. This can be
understood from the fact that the uplifting term goes like
$V_{uplift}\sim \frac{1}{a^2 Z^3}\epsilon^3$ which cannot compete
with the negative term in the moduli potential scaling as
$\sqrt{t}a^2 Z^2\epsilon^3$, and if one tries to increase the mass
parameter in order to compensate this the moduli vacuum starts to
be perturbed by sending their VEV to larger values so that the net
effect is lost. We have check numerically, in fact, that the
vacuum with a non-vanishing $m$, but still small, ends deeper than
in the $m=0$ case, then when increasing $m$ only a virtual uplift
is realized by the shrink of the potential due to a larger $T$
VEV.
\\
The $D$-term dynamics drives the leading contribution to the
$\phi$ real part mass, $m_{Re(\phi)}^2\sim(
\partial_\phi G_X)^2\sim  \phi^2/T^{2n}\sim \epsilon^{1+2n/3}$, so
canonically normalized $m_\phi \sim \sqrt{\epsilon}$, like the
gauge symmetry breaking scale. More precisely this real field has
a small component from the real part of $t$, and the imaginary
counterpart is the would-be Goldstone field eaten up by the now
massive vector field. The other combination of $\phi$ and $t$ gets
a mass like for the small modulus in the standard stabilization.
For the large modulus we have the same scaling for the mass as in
the standard scenario. For $\rho$ we read the mass from the
quadratic term eq.\eqref{rhocua}, which scales like
$m_\rho^2=\epsilon^{2+2n/3}$, so normalized $m_\rho\sim \epsilon$
for both real and imaginary components.

\subsubsection*{$N_C<N_F<3/2N_C$ case}\label{KQModel}

An interesting possibility was studied in
\cite{Krippendorf:2009zza}, where standing in the Seiberg duality
window $N_C<N_F<3/2N_C$ of the non-Abelian sector found vacua with
exponentielly sized volumina. The dual model is described by a
meson like field, $\phi$, and two chiral fields, $q$ and $p$, with
a superpotential depending on the non-perturbative scale,
$\Lambda$, a possible mass term for $\phi$ and a further scale
$\mu$ dictated by the duality relations.
As usual the non-perturbative scale has an exponential dependency
on the small modulus. Then replacing the mass parameter by a
dynamical field $\rho$, we have the following superpotential
\cite{Krippendorf:2009zza},
\begin{equation}
W=W_o+A\, e^{-a\,t}\left(\frac{q\phi p}{\mu}+\rho\phi\right)\,,
\end{equation}
which has an $U(1)$ symmetry characterized be the charges,
\begin{center}
\begin{tabular}{c|c|c|c|c|c}
&$q$&$p$&$\rho$&$\phi$&$e^{a\,t}$\\\hline $U(1)$ &
$1$&$1$&$2$&$-1$&$1$
\end{tabular}
\end{center}
so that $t$ gets a non-linear realization identifying
$\delta^t=2/a$ as the $GS$ coefficient for $t$.
\\
With this superpotential and particular modular weights,
$K\supset\frac{t^n}{T_r}|Q|^2$, Krippendorf and Quevedo found
vacua with exponentially large volumina and vanishing cosmological
constant \cite{Krippendorf:2009zza}. In the following we will give
a generic study of the scalings around the vacuum. To simplify the
analysis we neglect the fields $q$ and $p$, as their VEV turn out
to vanish \cite{Krippendorf:2009zza}.
\\
Taking for $K\supset\frac{Z(t_r)}{T_r^n} |Q|^2$, and $f_X=t$ the
leading contributions to the $D$-term potential reads
\begin{equation}
V_D\sim\frac{1}{t_r}\left(\frac{3}{2a}\frac{\sqrt{t_r}}{T_r^{3/2}}+\frac{Z(t_r)}{T_r^n}
\big(2|\rho|^2-|\phi|^2\big)\right)^2\,,
\end{equation}
neglecting also terms subleading in $1/at$. Like in the previous
two cases the minimization of the potential would lead to a
leading cancellation of the $D$-term potential due a non-vanishing
$\phi$ VEV, then $|\phi|^2\approx \frac{3\sqrt{t_r}}{2a
Z(t_r)T_r^{3/2-n}}\sim \epsilon^{1-2n/3}\phi_o(t_r)^2$ taking {\it
a priori} $\langle \rho\rangle\ll \langle \phi\rangle$ and
$\phi_o(t_r)$ encoding ${\cal O}(1)$ factors. Plugging back this
solution into the scalar potential one finds the following leading
potential for $\rho$, regarding real solutions so to forget the
$r$ indices,
\begin{align}
V_F\approx \frac{
\phi_o^2A^2e^{-2a\,t}}{\lambda^2Z(t)T^{\frac{9-4n}{2}}}+4a\,t\frac{W_oA\,e^{-a\,t}}{\lambda^2T^{\frac{15-2n}{4}}}
\phi_o\rho+\frac{3\,\hat \xi
W_o^2}{2\lambda^3T^{9/2}}\,,~~~
V_D\approx 4\frac{Z(t)^2}{t\,T^{2n}}\rho^4\,.
\end{align}
We have tacitly neglected a term quadratic in $\rho$ coming from
$e^GG^{\phi\bar\phi}|G_\phi|^2\supset T^{n-3}e^{-a\,t}\rho^2$ as
can be check at the end, using the resulting scalings, it is
indeed irrelevant. Then the minimization for $\rho$ leads to
\begin{equation}
\rho=-\frac{1}{(2\lambda)^{2/3}}\left(\frac{3\,a}{2}\right)^{1/6}\frac{t^{3/4}}{Z(t)^{5/6}}\frac{\big(-W_o
A e^{-a\,t} \big)^{1/3}}{T^{5(3-2n)/12}}\,,
\end{equation}
so around this solution there is a scalar potential for the moduli
from which their stabilization can be studied,
\begin{align}
V_{mod}\approx &\frac{3}{2} \frac{A^2 e^{-2 a\, t} \sqrt{t}}{
  a\,\lambda^2\,Z(t)^2 T^{9/2-2n}} -
 \frac{3}{2\lambda^{8/3}} \Big(\frac{9}{2}\Big)^{1/3} \frac{(a\,W_o^2)^{2/3}t^2}{Z(t)^{4/3}}\frac{ A^{4/3} e^{-4 a t/3} }{  T^{5-4n/3}}
 +
 \frac{ 3 W_o^2 \hat \xi }{2 \lambda^3 T^{9/2}}\,.
\end{align}
As was noticed in \cite{Krippendorf:2009zza} the structure of the
resulting potential is quite similar to the one of the standard
scenario though the powers of $T$ do not allow for a precise
analytical study. They find, however, numerically vacua with
exponentially large volume which furthermore realize vanishing
cosmological constant.
\\
Let us push a bit further the analysis for the stabilization, by
writing the potential for the moduli as $V_{mod}=\big(\tilde A
e^{-2a\,t}T^{2n}-\tilde B e^{-4a\,t/3}T^{4n/3-1/2}+\tilde
C\big)/T^{9/2}$. By requiring a vanishing cosmological constant
the e.o.m. for $T$ reads,\footnote{The same conclusions are
reached using the leading e.o.m. for $t$, regarding all three
factors in the potential to be of the same order.}
\begin{equation}
\partial_TV_{mod}\approx \Big(2n\tilde A
e^{-2a\,t}T^{2n}-(4n/3-1/2)\tilde B
e^{-4a\,t/3}T^{4n/3-1/2}\Big)/T^{11/2}=0\,,
\end{equation}
so we get that $e^{2a\,t/3}\sim T^{2n/3+1/2}$, telling us that the
scaling of the two first terms in the moduli potential is the
same, more precisely ${\cal O}(T^{-14/3})={\cal
O}(\epsilon^{28/9})$. But if the cosmological constant is to be
zero the third term cannot have a larger scaling so one needs to
require a tuning in the parameters, for example requiring $\tilde
C\sim T^{-1/6}\sim \epsilon^{1/9}$. Such requirement is in fact
found in \cite{Krippendorf:2009zza} by doing an approximate
analytic study and looking for conditions for the solution to
exist. In their approach they consider the possibility of changing
a bit the powers of $T$ appearing in the potential, to be precise
by $T^{1/6}$, something valid as far the volume is not huge. In
this regime, then, we can neglect a possible tuning in $\tilde C$
or in any other parameter appearing in the potential and regard
them as ${\cal O}(1)$.
\\
Using the full e.o.m. for $\phi$ we estimate the VEV for the
$D$-term finding like in the previous cases $D^X\sim
G_X\sim\epsilon^2$. Unfortunately this rough analytic approach is
not fine enough so to provide a good starting point to look for
numerical solutions, then we have been not able to check these
properties numerically.

\section{Numerical tests}\label{NumTest}
This section shows numerical tests for the decoupling in the
models discussed analytically in the main text but for the
Krippendorf-Quevedo model \cite{Krippendorf:2009zza} (see section
\eqref{KQModel}), for which unfortunately so far we have not
managed to find numerically such vacua. Although the form of the
chosen K\"ahler potential and superpotential on each case are
string inspired, we sacrifice a precise justification of the
numerical parameters in favor of avoiding extra factors that can
make less clear the analysis of the results.

\subsection*{$H$-sector Toy-model}
In order to implement a numerical test we should choose a $H$
sector, which better to be as realistic as possible. Our main
assumption in the analysis was that the conditions
$\partial_HW_{SC}+W_{SC}\partial_H K_{SC}=0$ do not leave flat
directions in the $H$ fields. From the way the Dilaton enters in
the flux induced superpotential of type-IIB superstrings
\cite{Gukov:1999ya}, we need at least one more dynamical field in
order to satisfy such requirement.\footnote{Fixing $S$ alone is
still possible if one allows a constant part in the
superpotential, something that obviously we want to avoid.} In the
following examples we will take the minimal set of one Complex
structure field plus the Dilaton described by the K\"ahler
potential ${\cal K}_{cs}=-\log(S+\bar S)-\log(U+\bar U)$. The
superpotential, inspired from explicit calculations on flux
compactification \cite{Lust:2005dy}, is given by
\begin{equation}
W_{cs}(U,S)= a_{cs} U^2+b_{cs}U+S\big(f_{cs} U^2+d_{cs}
U+e_{cs}\big)\,.
\end{equation}
With numerical parameters that in all following examples we will
take as $a_{cs}=1$, $b_{cs}=-\frac12$, $d_{cs}=-3$, $f_{cs}=1$ and
$e_{cs}=5/2$, so that lead to the fixed values, solutions to the
$F$-flatness conditions, $S_o=U_o=1$, and more over
$W_{cs}(U_o,S_o)=W_o=1$.
\\
The other common feature in the following examples is the matter
independent K\"ahler potential, given by the one of type-IIB
compactifications with orientifolds and the leading
$\alpha^\prime$ corrections, eq.\eqref{ModKPot}, with the CY
volume \eqref{CP4}. Notice that all these features are fairly
natural and generic, and no particular intention drives its
election.

\subsection{Charged large modulus}
For these models we will take the following particular choice,
\begin{equation}
K=K_{mod}+\frac{1}{(U+\bar U)^\eta {\cal
V}^n}|Q|^2\,,~~W=W_{cs}-A\, U\, e^{-a\, t}\,,
\end{equation}
where we have introduced the mixing between the complex structure
moduli and the matter fields in the K\"ahler potential, and a
further coupling in the superpotential with the small modulus. The
Killing vectors for the systems, with ordering $\{U,S,T,t,\phi\}$,
are $\{0,0,i\delta^T/2,0,i q_Q Q\}$. The gauge kinetic function is
chosen to be $f_X=T+\kappa S$, $\kappa$ a real number.

\subsubsection*{Vanishing $Q$ VEV}
If the GS coefficient is taken as positive then a negative charge
for $Q$ leads to the vanishing VEV case presented in appendix
\ref{VanishingQVEVmodel} and whose decoupling was studied in
section \ref{TChargedSect}. We take the following values for the
parameters, besides the ones already fixed in the $W_{cs}$
superpotential,
\begin{equation}
\lambda=1\,,~~A=1\,,~a=5\pi\,,~\xi=2\,,~\kappa=1\,,~\eta=1\,,~n=\frac23\,,~q_Q=-1\,,~\delta^T=1\cdot
10^{-11}\,.
\end{equation}
The value for the GS coefficient $\delta^T$ is chosen such to fine
tune the cosmological constant. Table \ref{resultsQzero} shows the
solution for the vacuum and the deviation from the simplified
model where the $S$ and $U$ superfields are fixed to $So=U_o=1$,
with vanishing spinor and auxiliary components. The canonical
normalized auxiliary fields VEV, $F^I_c=|G_{I\bar I}F^I F^{\bar
I}|^{1/2}$ no sum, are also shown. The system corresponds to an
$\epsilon$ parameter, as defined in the main text, of roughly
${\cal O}(10^{-21})$, expecting then corrections to the simplified
model of this order.
\begin{table}[t]
\begin{center}
\begin{tabular}{|c|c|c|c|p{1.3cm}|}
\hline& $\langle X \rangle $ & $\Delta \langle X \rangle$ &$F^X_c$
&~~$\Delta F^X_c$ \\ \hline
$U$ & $1-4\cdot 10^{-20}$ & $-4\cdot 10^{-20}$   & $1\cdot 10^{-41}$ & ~~~~-- \\
$S$ & $1-1\cdot 10^{-19}$ &   $-1\cdot 10^{-19}$ & $3\cdot 10^{-41}$ & ~~~~-- \\
$T$ & $7.5\cdot 10^{13}$ &   $2\cdot 10^{-18}$  & $1\cdot 10^{-21}$ &  $3\cdot 10^{-18}$  \\
$t$ &  $3.21$             &  $7\cdot 10^{-20}$   & $1\cdot 10^{-33}$ & $5\cdot 10^{-18}$\\
\hline
\end{tabular}
\end{center}
\caption{VEV's of the fields and canonical normalized $F$-term,
and their relative shifts compared to the simplified model. The
$Q$ sector is not reported given the triviality of the results.
Here and in the main text $\Delta X \equiv (X-X_{sim})/X$. All
quantities are in natural units.}\label{resultsQzero}
\end{table}
The physical mass spectrum in $GeV$ units is
\begin{align}
&m_{t_i} =18.5\cdot 10^{-2}\,,~~&m_{t_r}= 18.4\cdot
10^{-2}\,,~~&m_{Q} =8.8\cdot 10^{-2}\,,~~ &m_{T_r}=7.0\cdot
10^{-14}\,,\cr &m_{\tilde S_r} = 3.4\cdot 10^{-2}\,, ~~&m_{\tilde
S_i} = 3.0\cdot 10^{-2}\,,~~&m_{\tilde U_i}= 4.5\cdot
10^{-3}\,,~~&m_{\tilde U_r} = 8.0\cdot 10^{-4}\,.\nn
\end{align}
where $\tilde S$ and $\tilde U$ are defined by a sum of the
original $S$ and $U$, but with main component $S$ and $U$
respectively, and the subscript $r$ ($i$) denote the real
(imaginary) field components. The imaginary component of $T$ is
the would-be Goldstone boson so is exactly massless.
\\
The shifts compared to the simplified model are
\begin{eqnarray}
\Delta m_{t} = 7\cdot 10^{-18}\,, \ \ \Delta m_{Q} = 5\cdot
10^{-18}\,, \ \ \Delta m_{T_r}= 8 \cdot 10^{-18}\,,
\end{eqnarray}
defining $\Delta X  \equiv (X-X_{sim})/X$. For the gravitino mass,
the $D$-auxiliary field VEV and the cosmological constant we have,
all in natural units,
\begin{align}
D^X&=1.3\cdot 10^{-39}\,,& \Delta D^X &=5\cdot 10^{-18} \,,&\nn \\
m_{3/2}&=  7.7\cdot 10^{-22} \,,&  \Delta m_{3/2}^2 &=3\cdot 10^{-18}\,,&  \\
V_0 &=  -5.5\cdot 10^{-65}\,,&  \Delta V_0 &= 9\cdot
10^{-18}\,.&\nn
\end{align}
Although a quite unrealistic scenario since presents a SUSY
breaking scale in the $10^{-3}GeV$ region and a too light moduli,
such small $\epsilon$ helps to split the scales and clearly shows
the scaling of the errors in the simplified version. Any way we
have check for different values of $\epsilon$ these scaling. From
the results we see that the mistake are few orders of magnitude
greater than the predicted values, reaching even three orders in
worst cases. Still, it is clear that all errors are correlated and
of the same order. In order to understand such difference with the
prediction we should remember the numerical factors carefully
extracted in section \ref{ONLYCHIRALVSsect}. There we found that
the error was enhanced by a factor $\sim \frac{8}{3}\lambda
t^{3/2} a^4 \tilde A^2$ with $\tilde A$ a numerical factor missed
in the analytical solution for $e^{at}$. This factor accounts for
an increasing of two orders in the result. The extra order is
easily understood from the fact that the error in the small
modulus propagates in the e.o.m. for the large one with a factor
$a$. Indeed the larger error in the fields VEV corresponds to $T$,
as can be seen in table \ref{resultsQzero}, and all the following
samples, and then it propagates to the auxiliary fields VEV's,
masses and more drastically in the cosmological constant, being
the first two suppressed by $e^{G/2}\sim 1/T^{3/2}$ and the last
one by $e^G\sim 1/T^3$, so that roughly $\Delta V\sim -3 V \Delta
T$.

\subsubsection*{Non-vanishing $Q$ VEV}

With the opposite sign for the $Q$ charge the VEV of $Q$ is non-
zero and an $F$-term uplifting is possible, section
\ref{nonzeroQVEVmodel}. The parameters taken for this example are
\begin{equation}
\lambda=1\,,~~A=1\,,~a=3\pi\,,~\xi=2\,,~\kappa=1\,,~\eta=1\,,~n=\frac49\,,~q_Q=1\,,~\delta^T=5\cdot
10^{-6}\,.
\end{equation}
The value for the GS coefficient $\delta^T$ again is chosen such
to tune the cosmological constant. Table \ref{resultsQnonzero}
shows the VEV's of the fields and their deviations from the
simplified model. The system corresponds to an $\epsilon$, as
defined in the main text, of roughly ${\cal O}(10^{-12})$,
expecting corrections to the simplified version of this order.
\begin{table}[t]
\begin{center}
\begin{tabular}{|c|c|c|c|p{1.3cm}|}
\hline& $\langle X \rangle $ & $\Delta \langle X \rangle$ &$F^X_c$
&~~$\Delta F^X_c$ \\ \hline
$U$ & $1-1\cdot 10^{-11}$   & $-1\cdot 10^{-11}$ & $1\cdot 10^{-24}$ &~~~~ --  \\
$S$ & $1-5\cdot 10^{-11}$   & $-5\cdot 10^{-11}$ & $3\cdot 10^{-24}$ &~~~~ -- \\
$T$ & $1.59\cdot 10^8$      & $-4\cdot 10^{-10}$ & $4\cdot 10^{-13}$ &  $6\cdot 10^{-10}$  \\
$t$ &  $3.21$               & $-2\cdot 10^{-11}$ & $1\cdot 10^{-20}$ &  $8\cdot 10^{-10}$ \\
$Q$ &  $1.2\cdot 10^{-4}$   & $6\cdot 10^{-11}$ & $1\cdot 10^{-20}$ &  $8\cdot 10^{-10}$ \\
 \hline
\end{tabular}
\end{center}\caption{VEV's of the
fields and canonical normalized $F$-term, and their relative
shifts compared to the simplified model with $\Delta X \equiv
(X-X_{sim})/X$. All quantities are in natural
units.}\label{resultsQnonzero}
\end{table}
The physical mass spectrum in $GeV$ units is
\begin{align}
&m_{Q_r} =4.2\cdot 10^{7}\,,~~&m_{t}= 3.6\cdot 10^{7}\,,~~&m_{Q_i}
=9.8\cdot 10^{5}\,,~~ &m_{T_r}=0.6\phantom{\cdot 10^{6}}\,,\cr
&m_{\tilde U_r}= 1.1\cdot 10^{7}\,,~~&m_{\tilde U_i} = 9.9\cdot
10^{6}\,,~~&m_{\tilde S_i} = 1.5\cdot 10^{6}\,, ~~&m_{\tilde S_r}
= 2.6\cdot 10^{5}\,.
\end{align}
again $\tilde S$ and $\tilde U$ are defined by a sum of the
original $S$ and $U$, but with main component $S$ and $U$
respectively and the subscript $r$ ($i$) denote the real
(imaginary) field components. The imaginary component of $T$ is
the would-be Goldstone boson so is exactly massless.
\\
The shifts compared to the simplified model are
\begin{eqnarray}
\Delta m_{t} = 5\cdot 10^{-10}\,, \ \ \Delta m_{Q_r} = 4\cdot
10^{-10}\,,\ \ \Delta m_{Q_i} = 6 \cdot 10^{-10}\,, \ \ \Delta
m_{T_r}= 6 \cdot 10^{-10}\,.
\end{eqnarray}
For the gravitino mass, the $D$-auxiliary field VEV and the
cosmological constant, in natural units, we have
\begin{align}
D^X&=2.1\cdot 10^{-26}\,,& \Delta D^X &=1\cdot 10^{-9} \,,&\nn \\
m_{3/2}&=  2.5\cdot 10^{-13} \,,&  \Delta m_{3/2}^2 &=6\cdot 10^{-10}\,,&  \\
V_0 &=  -4.3\cdot 10^{-39}\,,&  \Delta V_0 &= 1\cdot
10^{-9}\,.&\nn
\end{align}
This, now more realistic, scenario, with a SUSY breaking scale of
${\cal O}(10^6)GeV$ presents exactly the same features in the
errors as the previous case. This is expected as the largest
corrections come from the same source independent of the matter
fields and gauge dynamics. Thus, the explanation for the factors
showing the errors larger than expected ones follows verbatim
here.
\\
For the record we have done numerically the very same model
changing the gauge kinetic function to $f=t+\kappa S$. This in
order to check the claim that is the scaling of the VEV for $D^X$
auxiliary field what matters. Otherwise, as explained in footnote
\ref{importantFN}, with $f\sim t$ the corrections would rather
scale like ${\cal O}(\epsilon^{2/3})$. We find in fact numerically
the very same results as the ones just presented, being the only
drastic change the VEV of $G_X$.

\subsection{Charged small modulus}\label{appNewModelNum}
The system is the one described in detail in appendix
\eqref{appNewModel}, so besides checking explicitly in numbers the
decoupling we take the chance to see all features found
analytically there. For the matter fields the K\"ahler potential
is taken to be
\begin{equation}
K_Q=\frac{t_r}{T_r}\left(|\rho|^2+U_r|\phi|^2\right)\,.
\end{equation}
The superpotential is taken as
\begin{equation}
W=W_{cs}(U,S)-A\,S\frac{e^{-a\,t}}{\phi^2}-m\,U^2\, \rho\,
\phi^2\,,
\end{equation}
and the gauge kinetic function $f_X=t+\kappa S$, which is
different from the one taken in \eqref{appNewModel} but its $t$
dependency and scaling in $\epsilon$ are the same though. The
parameters taken are
\begin{equation}
\lambda=1\,,~~A=1\,,~a=3\pi\,,~\xi=\frac{6}{5}\,,~\kappa=1\,,~m=268\cdot
10^{-5}\,,~\delta^t=\frac{2}{3\pi}\,,q_\phi=-\frac12\,,q_\rho=1\,.
\end{equation}
We take a value for $m$ that in principle would work to uplift the
vacuum, being ${\cal O}(\epsilon^{1/2})$. However the relative
uplift to the zero $m$ vacuum turn to be only a virtual one being
only due to the shift caused on the VEV of $T$ which moves to
larger values. In fact, as advertised before, for $m$ slightly
smaller than $\epsilon^{1/2}$, such that its perturbation on $T$
is less relevant the vacuum goes slightly deeper than in the zero
$m$ case, the same happening with positive or negative values for
$m$. As defined in appendix \eqref{appNewModel} the GS coefficient
is related to the parameter $a$ by $\delta^t=2/a$, and the $\phi$
and $\rho$ charges are fixed as well, but we report them again for
completeness.
\\
Table \ref{resultsNFlNc} shows VEV's of the fields and the
deviation from the simplified model. The canonical normalized
auxiliary fields VEV are also shown. The system corresponds to an
$\epsilon$, as defined in the main text, of roughly ${\cal
O}(10^{-6})$, to be precise $8\cdot 10^{-7}$, expecting
corrections to the simplified model of order ${\cal O}
(\epsilon^{2/3})\sim 10^{-5}$.
\begin{table}[t]
\begin{center}
\begin{tabular}{|c|c|c|c|p{1.3cm}|}
\hline& $\langle X \rangle $ & $\Delta \langle X \rangle$ &$F^X_c$
&~~$\Delta F^X_c$ \\ \hline
$U$   & $1-1\cdot 10^{-5}$   & $-1\cdot 10^{-5}$    & $2\cdot 10^{-12}$ &~~~~ --  \\
$S$   & $1-4\cdot 10^{-5}$   & $-4\cdot 10^{-5}$    & $5\cdot 10^{-12}$ &~~~~ -- \\
$T$   & $11247$              & $-2\cdot 10^{-4}$    & $7\cdot 10^{-7}$  &  $3\cdot 10^{-4}$  \\
$t$   &  $2.36$              & $-2\cdot 10^{-5}$    & $4\cdot 10^{-11}$ &  $4\cdot 10^{-4}$ \\
$\phi$ & $4.6\cdot 10^{-2}$  &  $7\cdot 10^{-5}$    & $9\cdot 10^{-11}$ &  $4\cdot 10^{-4}$ \\
$\rho$ & $7.1\cdot 10^{-3} $ &  $8\cdot 10^{-5}$   & $2\cdot 10^{-10}$ &  $3\cdot 10^{-4}$ \\
\hline
\end{tabular}
\end{center}
\caption{VEV's of the fields and canonical normalized $F$-terms,
and their relative shifts, derived by a numerical analysis with
$\Delta X \equiv (X-X_{sim})/X$. All quantities are in natural
units.}\label{resultsNFlNc}
\end{table}
We check then that indeed the field $\phi$ is
stabilized mainly by the nearly vanishing $D$-term condition,
$\phi\approx(a^2t_r T_r)^{-1/4}$. For the rest of the quantities
the scalings are also in agreement with the one found in appendix
\eqref{appNewModel}. The physical mass spectrum in $GeV$ is
\begin{align}
&m_{\tilde \phi_r} = 6.3\cdot 10^{14}\,,&~m_{\tilde t} =4.6\cdot
10^{13}\,,&~~~m_{\rho_i} =2.2\cdot 10^{13}\,,&~m_{\rho_r}=2.0\cdot
10^{12}\,,\cr &m_{\tilde U_r} = 1.9\cdot 10^{13}\,,&m_{\tilde U_i}
= 1.7\cdot 10^{13}\,,&~~~m_{\tilde S_i}=
2.5\cdot 10^{12}\,,&m_{\tilde S_r}=4.4\cdot 10^{11}\,,\\
&m_{T_r}=1.2\cdot 10^{9}\,.&\phantom{m_{\tilde U_r} = 3.1\cdot
10\,,}&\phantom{m_{\tilde U_r} = 3.1\cdot
10\,,}&\phantom{m_{\tilde U_r} = 3.1\cdot 10\,,}\nn
\end{align}
where $\tilde \phi$ ($\tilde t$) is a linear combination of $\phi$
and $t$ with $\phi$ ($t$) as main component, and $\tilde S$ and
$\tilde U$ are defined by a sum of the original $S$ and $U$, but
with main component $S$ and $U$ respectively. The subscript $r$
($i$) denote real (imaginary) field components. The imaginary
component of $\tilde \phi$ is the would-be Goldstone boson so is
exactly massless, as well the imaginary part  of $T$.
\\
The shifts compared to the simplified model are
\begin{eqnarray}
\Delta m_{\tilde\phi_r} = 2\cdot 10^{-4}\,, \ \ \Delta m_{\tilde
t} = 3\cdot 10^{-4}\,,\ \ \Delta m_{\rho} = 2\cdot 10^{-4}\,, \ \
\Delta m_{T_r}= 3 \cdot 10^{-4}\,.
\end{eqnarray}
For the gravitino mass, the $D$-auxiliary field VEV and the
cosmological constant we have, in natural units,
\begin{align}
D^X&=4.5\cdot 10^{-13}\,,& \Delta D^X &=5\cdot 10^{-4} \,,&\nn \\
m_{3/2}&=  4.2\cdot 10^{-7}\,,&  \Delta m_{3/2}^2 &=3\cdot 10^{-4}\,,&  \\
V_0 &=  -1.7\cdot 10^{-20}\,,&  \Delta V_0 &= 7\cdot
10^{-4}\,.&\nn
\end{align}
This scenario with high SUSY breaking scale, ${\cal
O}(10^{12})GeV$, shows an interesting situation that might happen
with the corrections. Although the corrections fall in predicted
order, i.e., ${\cal O}(\epsilon^{2/3})$, the main corrections come
from the ${\cal O}(\epsilon)$ contribution coming again mainly
from the $G_{t\bar t}|F^t|^2$ terms. This is check by performing
the same numerical search but setting $m=0$. Analytically it can
be understood by looking to enhancement factors as we did before
to explain the numerical results for the previous two cases. For
the corrections coming from the mass parameter we have $\Delta\sim
\frac{a^2\sqrt{t} \tilde A \tilde m}{\tilde \rho}\epsilon^{2/3}$,
with $\tilde A\equiv e^{-a\,t}/\epsilon^{4/3}$, $\tilde m\equiv
m/\epsilon^{1/2}$, $\tilde \rho\equiv \rho/\epsilon^{1/2}$, and
for the present example the correction factor is $\sim 1/2$. On
the other hand, the ones from $G_{it}\sim a
e^{-at}/\phi^2+\epsilon^{2/3}\phi^2$ leads to corrections going
like $\Delta\sim a^4 t^{3/2} \tilde A^2 \epsilon$, which enhance
them by almost two orders. The result is that both errors enter in
the same order though the ones scaling as ${\cal O}(\epsilon)$
turn out to be slightly bigger.

\bibliographystyle{JHEP}

\begin{thebibliography}{10}

\bibitem{Appelquist:1974tg}
T.~Appelquist and J.~Carazzone, {\it {Infrared Singularities and Massive
  Fields}},  {\em Phys. Rev.} {\bf D11} (1975) 2856.

\bibitem{Witten:1985xb}
E.~Witten, {\it {Dimensional Reduction of Superstring Models}},  {\em Phys.
  Lett.} {\bf B155} (1985) 151.

\bibitem{Giddings:2001yu}
S.~B. Giddings, S.~Kachru, and J.~Polchinski, {\it {Hierarchies from fluxes in
  string compactifications}},  {\em Phys. Rev.} {\bf D66} (2002) 106006,
  [\href{http://xxx.lanl.gov/abs/hep-th/0105097}{{\tt hep-th/0105097}}].

\bibitem{deAlwis:2005tf}
S.~P. de~Alwis, {\it {Effective potentials for light moduli}},  {\em Phys.
  Lett.} {\bf B626} (2005) 223--229,
  [\href{http://xxx.lanl.gov/abs/hep-th/0506266}{{\tt hep-th/0506266}}].

\bibitem{deAlwis:2005tg}
S.~P. de~Alwis, {\it {On integrating out heavy fields in SUSY theories}},  {\em
  Phys. Lett.} {\bf B628} (2005) 183--187,
  [\href{http://xxx.lanl.gov/abs/hep-th/0506267}{{\tt hep-th/0506267}}].

\bibitem{Achucarro:2007qa}
A.~Achucarro and K.~Sousa, {\it {F-term uplifting and moduli stabilization
  consistent with Kahler invariance}},  {\em JHEP} {\bf 0803} (2008) 002,
  [\href{http://xxx.lanl.gov/abs/0712.3460}{{\tt arXiv:0712.3460}}].

\bibitem{Choi:2008hn}
K.~Choi, K.~S. Jeong, and K.-I. Okumura, {\it {Flavor and CP conserving moduli
  mediated SUSY breaking in flux compactification}},  {\em JHEP} {\bf 07}
  (2008) 047, [\href{http://xxx.lanl.gov/abs/0804.4283}{{\tt
  arXiv:0804.4283}}].

\bibitem{Achucarro:2008sy}
A.~Achucarro, S.~Hardeman, and K.~Sousa, {\it {Consistent Decoupling of Heavy
  Scalars and Moduli in N=1 Supergravity}},  {\em Phys. Rev.} {\bf D78} (2008)
  101901, [\href{http://xxx.lanl.gov/abs/0806.4364}{{\tt arXiv:0806.4364}}].

\bibitem{Achucarro:2008fk}
A.~Achucarro, S.~Hardeman, and K.~Sousa, {\it {F-term uplifting and the
  supersymmetric integration of heavy moduli}},  {\em JHEP} {\bf 11} (2008)
  003, [\href{http://xxx.lanl.gov/abs/0809.1441}{{\tt arXiv:0809.1441}}].

\bibitem{Gallego:2008qi}
D.~Gallego and M.~Serone, {\it {An Effective Description of the Landscape -
  I}},  {\em JHEP} {\bf 01} (2009) 056,
  [\href{http://xxx.lanl.gov/abs/0812.0369}{{\tt arXiv:0812.0369}}].

\bibitem{Choi:2009jn}
K.~Choi, K.~S. Jeong, S.~Nakamura, K.-I. Okumura, and M.~Yamaguchi, {\it
  {Sparticle masses in deflected mirage mediation}},  {\em JHEP} {\bf 04}
  (2009) 107, [\href{http://xxx.lanl.gov/abs/0901.0052}{{\tt
  arXiv:0901.0052}}].

\bibitem{Gallego:2009px}
D.~Gallego and M.~Serone, {\it {An Effective Description of the Landscape -
  II}},  {\em JHEP} {\bf 06} (2009) 057,
  [\href{http://xxx.lanl.gov/abs/0904.2537}{{\tt arXiv:0904.2537}}].

\bibitem{Brizi:2009nn}
L.~Brizi, M.~Gomez-Reino, and C.~A. Scrucca, {\it {Globally and locally
  supersymmetric effective theories for light fields}},  {\em Nucl. Phys.} {\bf
  B820} (2009) 193--212, [\href{http://xxx.lanl.gov/abs/0904.0370}{{\tt
  arXiv:0904.0370}}].

\bibitem{Achucarro:2010da}
A.~Achucarro, J.-O. Gong, S.~Hardeman, G.~A. Palma, and S.~P. Patil, {\it
  {Features of heavy physics in the CMB power spectrum}},  {\em JCAP} {\bf
  1101} (2011) 030, [\href{http://xxx.lanl.gov/abs/1010.3693}{{\tt
  arXiv:1010.3693}}].

\bibitem{Brizi:2010ab}
L.~Brizi and C.~A. Scrucca, {\it {Effects of heavy modes on vacuum stability in
  supersymmetric theories}},  {\em JHEP} {\bf 11} (2010) 134,
  [\href{http://xxx.lanl.gov/abs/1009.0668}{{\tt arXiv:1009.0668}}].

\bibitem{Achucarro:2010jv}
A.~Achucarro, J.-O. Gong, S.~Hardeman, G.~A. Palma, and S.~P. Patil, {\it {Mass
  hierarchies and non-decoupling in multi-scalar field dynamics}},
  \href{http://xxx.lanl.gov/abs/1005.3848}{{\tt arXiv:1005.3848}}.

\bibitem{Kachru:2003aw}
S.~Kachru, R.~Kallosh, A.~Linde, and S.~P. Trivedi, {\it {De Sitter vacua in
  string theory}},  {\em Phys. Rev.} {\bf D68} (2003) 046005,
  [\href{http://xxx.lanl.gov/abs/hep-th/0301240}{{\tt hep-th/0301240}}].

\bibitem{Balasubramanian:2005zx}
V.~Balasubramanian, P.~Berglund, J.~P. Conlon, and F.~Quevedo, {\it
  {Systematics of Moduli Stabilisation in Calabi-Yau Flux Compactifications}},
  {\em JHEP} {\bf 03} (2005) 007,
  [\href{http://xxx.lanl.gov/abs/hep-th/0502058}{{\tt hep-th/0502058}}].

\bibitem{Conlon:2005ki}
J.~P. Conlon, F.~Quevedo, and K.~Suruliz, {\it {Large-volume flux
  compactifications: Moduli spectrum and D3/D7 soft supersymmetry breaking}},
  {\em JHEP} {\bf 08} (2005) 007,
  [\href{http://xxx.lanl.gov/abs/hep-th/0505076}{{\tt hep-th/0505076}}].

\bibitem{Conlon:2006wz}
J.~P. Conlon, S.~S. Abdussalam, F.~Quevedo, and K.~Suruliz, {\it {Soft SUSY
  breaking terms for chiral matter in IIB string compactifications}},  {\em
  JHEP} {\bf 01} (2007) 032,
  [\href{http://xxx.lanl.gov/abs/hep-th/0610129}{{\tt hep-th/0610129}}].

\bibitem{Cremades:2007ig}
D.~Cremades, M.~P. Garcia~del Moral, F.~Quevedo, and K.~Suruliz, {\it {Moduli
  stabilisation and de Sitter string vacua from magnetised D7 branes}},  {\em
  JHEP} {\bf 05} (2007) 100,
  [\href{http://xxx.lanl.gov/abs/hep-th/0701154}{{\tt hep-th/0701154}}].

\bibitem{Conlon:2008cj}
J.~P. Conlon, R.~Kallosh, A.~D. Linde, and F.~Quevedo, {\it {Volume Modulus
  Inflation and the Gravitino Mass Problem}},  {\em JCAP} {\bf 0809} (2008)
  011, [\href{http://xxx.lanl.gov/abs/0806.0809}{{\tt arXiv:0806.0809}}].

\bibitem{Cicoli:2008va}
M.~Cicoli, J.~P. Conlon, and F.~Quevedo, {\it {General Analysis of LARGE Volume
  Scenarios with String Loop Moduli Stabilisation}},  {\em JHEP} {\bf 10}
  (2008) 105, [\href{http://xxx.lanl.gov/abs/0805.1029}{{\tt
  arXiv:0805.1029}}].

\bibitem{Conlon:2008wa}
J.~P. Conlon, A.~Maharana, and F.~Quevedo, {\it {Towards Realistic String
  Vacua}},  {\em JHEP} {\bf 05} (2009) 109,
  [\href{http://xxx.lanl.gov/abs/0810.5660}{{\tt arXiv:0810.5660}}].

\bibitem{Anguelova:2009ht}
L.~Anguelova, V.~Calo, and M.~Cicoli, {\it {LARGE Volume String
  Compactifications at Finite Temperature}},  {\em JCAP} {\bf 0910} (2009) 025,
  [\href{http://xxx.lanl.gov/abs/0904.0051}{{\tt arXiv:0904.0051}}].

\bibitem{Conlon:2010ji}
J.~P. Conlon and F.~G. Pedro, {\it {Moduli Redefinitions and Moduli
  Stabilisation}},  {\em JHEP} {\bf 1006} (2010) 082,
  [\href{http://xxx.lanl.gov/abs/1003.0388}{{\tt arXiv:1003.0388}}].

\bibitem{Conlon:2010jq}
J.~P. Conlon and F.~G. Pedro, {\it {Moduli-Induced Vacuum Destabilisation}},
  {\em JHEP} {\bf 05} (2011) 079,
  [\href{http://xxx.lanl.gov/abs/1010.2665}{{\tt arXiv:1010.2665}}].

\bibitem{Cicoli:2010yj}
M.~Cicoli and A.~Mazumdar, {\it {Inflation in string theory: a graceful exit to
  the real world}},  {\em Phys. Rev.} {\bf D83} (2011) 063527,
  [\href{http://xxx.lanl.gov/abs/1010.0941}{{\tt arXiv:1010.0941}}].

\bibitem{Krippendorf:2009zza}
S.~Krippendorf and F.~Quevedo, {\it {Metastable SUSY Breaking, de Sitter Moduli
  Stabilisation and K\'ahler Moduli Inflation}},  {\em JHEP} {\bf 11} (2009)
  039, [\href{http://xxx.lanl.gov/abs/0901.0683}{{\tt arXiv:0901.0683}}].

\bibitem{Binetruy:2004hh}
P.~Binetruy, G.~Dvali, R.~Kallosh, and A.~Van~Proeyen, {\it {Fayet-Iliopoulos
  terms in supergravity and cosmology}},  {\em Class. Quant. Grav.} {\bf 21}
  (2004) 3137--3170, [\href{http://xxx.lanl.gov/abs/hep-th/0402046}{{\tt
  hep-th/0402046}}].

\bibitem{Inprep}
D.~Gallego, {\it {Light Field Integration in SUGRA Theories}}, {\em in preparation}.

\bibitem{Conlon:2006tj}
J.~P. Conlon, D.~Cremades, and F.~Quevedo, {\it {Kaehler potentials of chiral
  matter fields for Calabi-Yau string compactifications}},  {\em JHEP} {\bf 01}
  (2007) 022, [\href{http://xxx.lanl.gov/abs/hep-th/0609180}{{\tt
  hep-th/0609180}}].

\bibitem{Kaku:1978nz}
M.~Kaku, P.~K. Townsend, and P.~van Nieuwenhuizen, {\it {Properties of
  Conformal Supergravity}},  {\em Phys. Rev.} {\bf D17} (1978) 3179.

\bibitem{Kugo:1982mr}
T.~Kugo and S.~Uehara, {\it {Improved Superconformal Gauge Condition in the N=1
  Supergravity Yang-Mills Matter System}},  {\em Nucl. Phys.} {\bf B222} (1983)
  125.

\bibitem{Kugo:1982cu}
T.~Kugo and S.~Uehara, {\it {Conformal and Poincare Tensor Calculi in N=1
  Supergravity}},  {\em Nucl. Phys.} {\bf B226} (1983) 49.

\bibitem{Ferrara:1983dh}
S.~Ferrara, L.~Girardello, T.~Kugo, and A.~Van~Proeyen, {\it {Relation Between
  Different Auxiliary Field Formulations of N=1 Supergravity Coupled to
  Matter}},  {\em Nucl. Phys.} {\bf B223} (1983) 191.

\bibitem{Choi:2004sx}
K.~Choi, A.~Falkowski, H.~P. Nilles, M.~Olechowski, and S.~Pokorski, {\it
  {Stability of flux compactifications and the pattern of supersymmetry
  breaking}},  {\em JHEP} {\bf 11} (2004) 076,
  [\href{http://xxx.lanl.gov/abs/hep-th/0411066}{{\tt hep-th/0411066}}].

\bibitem{Becker:2002nn}
K.~Becker, M.~Becker, M.~Haack, and J.~Louis, {\it {Supersymmetry breaking and
  alpha'-corrections to flux induced potentials}},  {\em JHEP} {\bf 06} (2002)
  060, [\href{http://xxx.lanl.gov/abs/hep-th/0204254}{{\tt hep-th/0204254}}].

\bibitem{Denef:2004dm}
F.~Denef, M.~R. Douglas, and B.~Florea, {\it {Building a better racetrack}},
  {\em JHEP} {\bf 06} (2004) 034,
  [\href{http://xxx.lanl.gov/abs/hep-th/0404257}{{\tt hep-th/0404257}}].

\bibitem{Gukov:1999ya}
S.~Gukov, C.~Vafa, and E.~Witten, {\it {CFT's from Calabi-Yau four-folds}},
  {\em Nucl. Phys.} {\bf B584} (2000) 69--108,
  [\href{http://xxx.lanl.gov/abs/hep-th/9906070}{{\tt hep-th/9906070}}].

\bibitem{Lust:2004fi}
D.~Lust, S.~Reffert, and S.~Stieberger, {\it {Flux-induced Soft Supersymmetry
  Breaking in Chiral Type IIB Orientifolds with D3/D7-Branes}},  {\em Nucl.
  Phys.} {\bf B706} (2005) 3--52,
  [\href{http://xxx.lanl.gov/abs/hep-th/0406092}{{\tt hep-th/0406092}}].

\bibitem{GomezReino:2007qi}
M.~Gomez-Reino and C.~A. Scrucca, {\it {Metastable supergravity vacua with F
  and D supersymmetry breaking}},  {\em JHEP} {\bf 08} (2007) 091,
  [\href{http://xxx.lanl.gov/abs/0706.2785}{{\tt arXiv:0706.2785}}].

\bibitem{Haack:2006cy}
M.~Haack, D.~Krefl, D.~Lust, A.~Van~Proeyen, and M.~Zagermann, {\it {Gaugino
  condensates and D-terms from D7-branes}},  {\em JHEP} {\bf 01} (2007) 078,
  [\href{http://xxx.lanl.gov/abs/hep-th/0609211}{{\tt hep-th/0609211}}].

\bibitem{Jockers:2005zy}
H.~Jockers and J.~Louis, {\it {D-terms and F-terms from D7-brane fluxes}},
  {\em Nucl. Phys.} {\bf B718} (2005) 203--246,
  [\href{http://xxx.lanl.gov/abs/hep-th/0502059}{{\tt hep-th/0502059}}].

\bibitem{Binetruy:1996uv}
P.~Binetruy and E.~Dudas, {\it {Gaugino condensation and the anomalous U(1)}},
  {\em Phys. Lett.} {\bf B389} (1996) 503--509,
  [\href{http://xxx.lanl.gov/abs/hep-th/9607172}{{\tt hep-th/9607172}}].

\bibitem{ArkaniHamed:1998nu}
N.~Arkani-Hamed, M.~Dine, and S.~P. Martin, {\it {Dynamical supersymmetry
  breaking in models with a Green- Schwarz mechanism}},  {\em Phys. Lett.} {\bf
  B431} (1998) 329--338, [\href{http://xxx.lanl.gov/abs/hep-ph/9803432}{{\tt
  hep-ph/9803432}}].

\bibitem{Dudas:2007nz}
E.~Dudas, Y.~Mambrini, S.~Pokorski, and A.~Romagnoni, {\it {Moduli
  stabilization with Fayet-Iliopoulos uplift}},  {\em JHEP} {\bf 04} (2008)
  015, [\href{http://xxx.lanl.gov/abs/0711.4934}{{\tt arXiv:0711.4934}}].

\bibitem{Gallego:2008sv}
D.~Gallego and M.~Serone, {\it {Moduli Stabilization in non-Supersymmetric
  Minkowski Vacua with Anomalous U(1) Symmetry}},  {\em JHEP} {\bf 08} (2008)
  025, [\href{http://xxx.lanl.gov/abs/0807.0190}{{\tt arXiv:0807.0190}}].

\bibitem{Dudas:2008qf}
E.~Dudas, Y.~Mambrini, S.~Pokorski, A.~Romagnoni, and M.~Trapletti, {\it {Gauge
  vs. Gravity mediation in models with anomalous U(1)'s}},  {\em JHEP} {\bf 03}
  (2009) 011, [\href{http://xxx.lanl.gov/abs/0809.5064}{{\tt
  arXiv:0809.5064}}].

\bibitem{Affleck:1983mk}
I.~Affleck, M.~Dine, and N.~Seiberg, {\it {Dynamical Supersymmetry Breaking in
  Supersymmetric QCD}},  {\em Nucl. Phys.} {\bf B241} (1984) 493--534.

\bibitem{Anguelova:2010qd}
L.~Anguelova and C.~Quigley, {\it {Quantum Corrections to Heterotic Moduli
  Potentials}},  {\em JHEP} {\bf 02} (2011) 113,
  [\href{http://xxx.lanl.gov/abs/1007.5047}{{\tt arXiv:1007.5047}}].

\bibitem{Lust:2005dy}
D.~Lust, S.~Reffert, W.~Schulgin, and S.~Stieberger, {\it {Moduli stabilization
  in type IIB orientifolds. I: Orbifold limits}},  {\em Nucl. Phys.} {\bf B766}
  (2007) 68--149, [\href{http://xxx.lanl.gov/abs/hep-th/0506090}{{\tt
  hep-th/0506090}}].

\end{thebibliography}
\providecommand{\href}[2]{#2}\begingroup\raggedright\endgroup

\end{document}